\documentclass[prx,reprint,amsmath,amssymb,aps]{revtex4-1}

\usepackage{bm}
\usepackage{graphicx}
\usepackage{booktabs}
\usepackage{slashed}
\usepackage{amssymb,amsmath,amsfonts}
\usepackage{color}
\usepackage[utf8]{inputenc}
\usepackage[caption=false]{subfig}
\usepackage{slashed}
\usepackage{dsfont}
\newcommand{\qq}{(Q)}
\newcommand{\Tr}{\mathrm{Tr}}

\newcommand{\tr}{\mathrm{tr}}

\def\eq#1{(\ref{#1})}

\def\Eq#1{Eq.~(\ref{#1})}
\def\Fig#1{Fig.~\ref{#1}}

\def\App#1{App.~\ref{#1}}

\definecolor{refcol}{RGB}{178,34,34}
\usepackage{microtype}
\usepackage[colorlinks,linkcolor=refcol,citecolor=refcol,urlcolor=refcol]{hyperref}

\definecolor{red}{rgb}{1,0,0}

\begin{document}

\title{The Myriad Uses of Instantons}

\author{Robert D. Pisarski}
\email{pisarski@bnl.gov}
\affiliation{Department of Physics, Brookhaven National Laboratory, Upton, NY 11973}

\author{Fabian Rennecke}
\email{frennecke@bnl.gov}
\affiliation{Department of Physics, Brookhaven National Laboratory, Upton, NY 11973}

\begin{abstract}
  In quantum chromodynamics (QCD), the role which topologically non-trivial configurations play in splitting the
  singlet pseudo-Goldstone meson, the $\eta^\prime$, from the octet is familiar.
  In addition, such configurations contribute
  to other processes which violate the axial $U(1)_A$ symmetry.
  While the nature of topological fluctuations in the confined phase is still unsettled, at temperatures well above that for
  the chiral phase transition, they can be described by a dilute gas of instantons.
  We show that instantons of arbitrary topological charge $Q$ generate anomalous interactions between $2 N_f |Q|$ quarks, 
  which for $Q = 1$ make the $\eta'$ heavy. 
  For two flavors we compute an anomalous quartic meson coupling 
  and discuss its implications for the phenomenology of the chiral phase transition.
  A dilute instanton gas suggests that for cold, dense quarks,
  instantons do not evaporate until {\it very} high densities, when the baryon chemical
  potential is $\gtrsim 2$~GeV.
\end{abstract}

\maketitle

In quantum chromodynamics (QCD), the up, down and strange quarks are relatively light, and
there is an approximate global flavor symmetry of $SU(3)_L \times SU(3)_R \times U(1)_A$.
When the hadronic vacuum spontaneously breaks chiral symmetry, a flavor octet of light pseudo-Goldstone bosons
is generated, which are the $\pi$, $K$, and $\eta$ mesons of broken $SU(3)_L \times SU(3)_R$.
When QCD first emerged, it was a puzzle why there isn't an associated ninth pseudo-Goldstone boson
in the flavor singlet channel, the $\eta'$, from the breaking of the axial $U(1)_A$ symmetry.

This occurs because while classically there is an axial $U(1)_A$ symmetry, it is not valid
quantum mechanically because of an anomaly \cite{Adler:1969gk, *Bell:1969ts,*Fujikawa:1979ay}.
There are topologically non-trivial fluctuations 
which violate the $U(1)_A$ symmetry \cite{Belavin:1975fg} and make the $\eta'$ heavy \cite{Fritzsch:1973pi}.
Classically these configurations are instantons: these have a topological winding number equal to an integer $Q$,
and an (Euclidean) action equal to $8 \pi^2 |Q| /g^2$, where $g$ is the coupling constant of QCD
\cite{tHooft:1976rip,*tHooft:1976snw,*tHooft:1986ooh,Callan:1976je,*Jackiw:1976pf,Callan:1976je,*Callan:1977gz,*Callan:1977qs,*Callan:1978bm,*Andrei:1978xg,Coleman:1978ae,tHooftunpub,*Witten:1976ck,*Jackiw:1976fs,Atiyah:1978ri,Corrigan:1977ma,*Christ:1978jy,*Bernard:1978ea,Grossman:1978,*Osborn:1978rn,*Corrigan:1978ce,*Corrigan:1979di,*Goddard:1980sw,*Osborn:1981yf,Gross:1980br,Altes:2014bwa,*Altes:2015wla,Shifman:1979uw,*Vainshtein:1981wh,*Shifman:1979nz,AragaodeCarvalho:1980de,*Baluni:1980db,*Chemtob:1980tu,Shuryak:1978yk,*Shuryak:1981ff,*Shuryak:1982dp,*Shuryak:1982hk,*Shuryak:1982qx,*Shuryak:1992ke,*Ilgenfritz:1988dh,*Ilgenfritz:1994nt,*Schafer:1995pz,*Schafer:1996wv,*Rapp:1997zu,*Rapp:1999qa,*Shuryak:2000df,*Schafer:2004gy,*Liao:2006ry,*Liao:2007mj,*Liu:2015ufa,*Liu:2015jsa,*Liu:2016mrk,*Liu:2016thw,*Liu:2016yij,*Liu:2018znq,*Lopez-Ruiz:2016bjl,*Lopez-Ruiz:2019oov,*Shuryak:2019zhv,Diakonov:1979nj,*Diakonov:1983hh,*Diakonov:1984vw,*Diakonov:1985eg,*Diakonov:1986tv,*Diakonov:2002fq,*Diakonov:2004jn,*Diakonov:2005qa,*Gromov:2005ij,*Diakonov:2007nv,*Bruckmann:2009nw,Diakonov:2009jq,Morris:1984zi,*Ringwald:1999ze,Dorey:2002ik,Schafer:1998up,*Son:2001jm,*Schafer:2002ty,*Schafer:2002yy,*Toublan:2005tn,*Hatsuda:2006ps,*Yamamoto:2008zw,*Chen:2009gv,*Yamamoto:2009ey,*Brauner:2009gu,*Abuki:2010jq,*Fukushima:2010bq,*Eto:2011mk,*Mitter:2013fxa,*Mitter:2013fxa,*S.:2019gpk,*Shim:2019yxn,Pawlowski:1996ch,*Gies:2002hq,*Gies:2006nz,*Braun:2018bik,*Braun:2019aow,Reinhardt:1988xu,*Osipov:2002wj,Jungnickel:1995fp,*Jungnickel:1996aa,*Grahl:2013pba,Dine:2010jg,*Dine:2014dga,*Dine:2017swf,Heller:2015box,*Fejos:2015xca,*Fejos:2016hbp,Dunne:2004sx,*Dunne:2005te,*Dunne:2005cy,Pisarski:2016ukx,Giacosa:2017pos,tHooft:1981nnx,*Sedlacek:1982cd,*vanBaal:1982ag,*Lee:1997vp,*Kraan:1998sn,*Kraan:1998pm,*Lee:1998bb,*Lee:1998vu,*vanBaal:2000zc,*Bruckmann:2002vy,*Bruckmann:2003ag,*Bruckmann:2004nu,*Poppitz:2008hr,Unsal:2007jx,*Shifman:2008ja,*Argyres:2012ka,*Poppitz:2012nz,*Basar:2013eka,*Dunne:2014bca,*Dunne:2016nmc,*Ogilvie:2012is,*Ogilvie:2014bwa,*Aitken:2018mbb,*Misumi:2019dwq,Hashimoto:2008xg,*Bruckmann:2009pa,*Ilgenfritz:2012gu,*Bornyakov:2013iva,*Bornyakov:2014esa,*Bornyakov:2015xao,*DiGiacomo:2015eva,*Frison:2016vuc,*Bornyakov:2017crk,*Itou:2018wkm,*Jahn:2018dke,*Giusti:2018cmp,Aoki:2012yj,*Cossu:2013uua,*Fukaya:2015ara,*Tomiya:2016jwr,*Aoki:2017paw,Bazavov:2012qja,*Buchoff:2013nra,*Dick:2015twa,*Petreczky:2016vrs,diCortona:2015ldu,*Bonati:2015vqz,*Borsanyi:2015cka,Brandt:2016daq,Vicari:2008jw,Bonati:2013tt}.
Instantons split the singlet $\eta'$ from the octet of pseudo-Goldstone bosons, and also
generate the $\theta$ parameter of QCD \cite{Callan:1976je,*Jackiw:1976pf}.

There are several open questions regarding the nature of topological fluctuations in the QCD vacuum. In absence of a large energy scale to cut-off the size of the instantons, their fluctuations on any length scale become relevant and the integration over their contribution blows up. This is cured non-perturbatively through confinement, where dense topologically non-trivial fluctuations may form an instanton liquid
\cite{Shuryak:1978yk,*Shuryak:1981ff,*Shuryak:1982dp,*Shuryak:1982hk,*Shuryak:1982qx,*Shuryak:1992ke,*Ilgenfritz:1988dh,*Ilgenfritz:1994nt,*Schafer:1995pz,*Rapp:1997zu,*Rapp:1999qa,*Shuryak:2000df,*Liu:2015ufa,*Liu:2015jsa,*Liu:2016mrk,*Liu:2016thw,*Liu:2016yij,*Liu:2018znq,Schafer:1996wv,Diakonov:1979nj,*Diakonov:1983hh,*Diakonov:1984vw,*Diakonov:1985eg,*Diakonov:1986tv,*Diakonov:2002fq,*Diakonov:2004jn,*Diakonov:2005qa,*Gromov:2005ij,*Diakonov:2007nv,*Bruckmann:2009nw,Diakonov:2009jq}.
Furthermore, it is expected that QCD behaves smoothly as the number of colors, $N_c$, goes to infinity
\cite{Witten:1978bc,*Witten:1979vv,*Witten:1980sp,*DAdda:1978dle,*DiVecchia:1980vpx,*DiVecchia:1980yfw,*DiVecchia:1980vpx, Lucini:2012gg,*Lucini:2013qja}.
In this limit, the contribution of a single instanton vanishes exponentially, 
while current algebra can be used to
show that the $\eta'$ is still split from the octet of pseudo-Goldstone bosons
\cite{Witten:1978bc,*Witten:1979vv,*Witten:1980sp,*DAdda:1978dle,*DiVecchia:1980vpx,*DiVecchia:1980yfw,*DiVecchia:1980vpx}.
This could occur if there are topologically
non-trivial fluctuations whose topological charge is not an integer, but an integer times $1/N_c$;
in certain limits, such as for adjoint QCD on a femto-torus, this can be shown semi-classically
\cite{tHooft:1981nnx,*Sedlacek:1982cd,*vanBaal:1982ag,*Lee:1997vp,*Kraan:1998sn,*Kraan:1998pm,*Lee:1998bb,*Lee:1998vu,*vanBaal:2000zc,*Bruckmann:2002vy,*Bruckmann:2003ag,*Bruckmann:2004nu,*Poppitz:2008hr,Unsal:2007jx,*Shifman:2008ja,*Argyres:2012ka,*Poppitz:2012nz,*Basar:2013eka,*Dunne:2014bca,*Dunne:2016nmc,*Ogilvie:2012is,*Ogilvie:2014bwa}.

However, if the effective coupling is small, {\it e.g.}\ at high temperature or quark density, then a semi-classical analysis is
valid, and topologically non-trivial fluctuations can be approximated as a dilute instanton gas
\cite{Gross:1980br,Altes:2014bwa,*Altes:2015wla}.
Numerical simulations of lattice QCD provide insight into how the topological structure changes
with temperature \cite{Aoki:2012yj,*Cossu:2013uua,*Fukaya:2015ara,*Tomiya:2016jwr,*Aoki:2017paw,Hashimoto:2008xg,*Bruckmann:2009pa,*Ilgenfritz:2012gu,*Bornyakov:2013iva,*Bornyakov:2014esa,*Bornyakov:2015xao,*DiGiacomo:2015eva,*Frison:2016vuc,*Bornyakov:2017crk,*Itou:2018wkm,*Jahn:2018dke,*Giusti:2018cmp,Bazavov:2012qja,*Buchoff:2013nra,*Dick:2015twa,*Petreczky:2016vrs,diCortona:2015ldu,*Bonati:2015vqz,*Borsanyi:2015cka,Brandt:2016daq,Vicari:2008jw,Bonati:2013tt}.
Remarkably, these demonstrate that the overall power of the topological susceptibility with respect to
the temperature $T$ {\it is} given by
a dilute instanton gas above temperatures as low as a few hundred MeV
\cite{Bazavov:2012qja,Buchoff:2013nra,diCortona:2015ldu,Bonati:2015vqz,Borsanyi:2015cka,Brandt:2016daq,Dick:2015twa,Petreczky:2016vrs}.

In this Letter we address a modest problem and consider quantities which are nonzero {\it only} because
of topologically nontrivial configurations, using a dilute instanton gas as an illustrative example.
Studies of the phenomenological implications of the axial anomaly, including the effects mentioned above, 
have been based on effective quark interactions that are generated in a dilute gas of instantons of unit topological charge \cite{tHooft:1976rip,*tHooft:1976snw,*tHooft:1986ooh}.
Here we generalize this by demonstrating that effective $2 N_f |Q|$-quark interactions are generated in a dilute gas of instantons of arbitrary topological charge $Q$ \cite{tHooftunpub,*Witten:1976ck,*Jackiw:1976fs, Atiyah:1978ri, Corrigan:1977ma,*Christ:1978jy}. Even though semi-classically such topological field configurations are suppressed exponentially, these interactions can give rise to novel anomalous effects related uniquely to fluctuations of higher topological charge. We explicitly work out the local effective interaction for $Q = \pm 2$ for the case where the constituent-instantons, which we define before \Eq{eq:ZQ}, are small.
At low energies and for two quark flavors this is a quartic meson interaction. We study its qualitative
impact on the mass spectrum within a simple mean-field picture. An appendix includes technical details of
the computation.

\textbf{Multi-instanton-induced interactions.}
We start with an analysis for arbitrary topological charge, generalizing that of 't Hooft \cite{tHooft:1976snw}. We consider the generating functional of QCD for Gaussian fluctuations around a background of instantons with topological charge $Q$,
which we term \emph{$Q$-instantons}. For a $Q$-(anti-) instanton background, massless quarks have $N_f |Q|$ (right-) left-handed zero modes \cite{Coleman:1978ae}. We show that the functional zero mode determinant of quarks has the structure of a $2 N_f |Q|$-quark correlation function and compute its coupling constant in a dilute gas of $Q$-instantons.

The zero modes of gauge fields arise from symmetries, such as translations, that yield inequivalent instanton solutions.
This defines a moduli space which is parametrized by the collective coordinates of the instantons.
The general $Q$-instanton has been constructed by Atiyah, Drinfeld, Hitchin and Manin (ADHM) \cite{Atiyah:1978ri, Christ:1978jy}. 
It can be viewed as a superposition of $Q$ individual instantons with unit charge, where each constituent is described by a location $z_i$, a size $\rho_i$ and an orientation in the gauge group $U_i$. There are then $4 N_c$ collective coordinates for each
constituent-instanton, so the moduli space of the $Q$-instanton has dimension $4 N_c |Q|$. Schematically,
the generating functional is
\begin{align}\label{eq:ZQ}
\begin{split}
Z^{\qq}[J] &= \int\!\!\mathcal{D}\chi \exp\Big\{\!-S[\chi + \chi^{\qq}] + \int_x \bar\psi J \psi \Big\}\\
&\approx \int\!\!dC_Q\, n_Q(C_Q) \det{}_{\!0} (J)\,,
\end{split}
\end{align}
where $\chi = (A_\mu, c, \bar c, \psi, \bar \psi)$ contains the fluctuating gluon, ghost and quark fields and $\chi^{\qq} = (A_\mu^{\qq}, 0, 0, 0, 0)$ contains the $Q$-instanton background field $A_\mu^{\qq}$;
to avoid notational clutter, the superscript $Q$ in $A_\mu^{\qq}$ denotes the topological charge.  $S[\chi]$ is the gauge-fixed action of QCD in Euclidean spacetime. 
In the second line we integrate the path integral over the non-zero modes to leading order in the saddle-point approximation, leaving only the integration over the collective coordinates $C_Q$. The instanton density $n_Q$ contains the functional determinants of the zero and non-zero modes of gluons and ghosts, the non-zero mode determinant of the quarks and the Jacobian from changing the integration over zero modes to collective coordinates \footnote{It is understood that all determinants are normalized with the corresponding determinant for a vanishing gluon background field.}.

Our main ingredient is the zero modes of massless quarks \cite{Corrigan:1978ce, Osborn:1978rn}. Due to the axial anomaly, the Dirac operator in the presence of the $Q$-instanton, $\slashed{D}^{\qq} = \gamma_\mu \big(\partial_\mu + A_\mu^{\qq}\big)$, has $N_f |Q|$ zero modes, $\slashed{D}^{\qq} \psi_{fi}^{\qq} = 0$, where $f = 1,\dots, N_f$ is an index for flavor and $i = 1, \dots, |Q|$ is a topological charge index.
Because of the zero modes, the generating functional is only nonzero in the presence of a source $J$, which generates
the quark zero mode determinant, $\det{}_{\!0} (J)$, in \Eq{eq:ZQ}.

The generating functional in \Eq{eq:ZQ} has first been computed for $Q = 1$ and $N_c = 2$ \cite{tHooft:1976rip} and arbitrary $N_c$ \cite{Bernard:1979qt}. For $|Q| >1$, the generating functional to one loop order is only known in certain limits \cite{Dorey:2002ik}.

One limit where one can compute is when
the distance between the locations of the constituent-instantons are much larger than their sizes; {\it i.e.}\ $|R_{ij}| \equiv |z_i - z_j| \gg \rho_i$ for all $i \neq j$. In this case, at leading order, the $Q$-instanton can be viewed as $|Q|$ instantons of unit charge which are well separated. Expanding the general ADHM-solution in this small limit \cite{Christ:1978jy}, the path integral factorizes into into a product of constituent-instanton contributions,
\begin{align}\label{eq:ZQdilute}
Z^{\qq}[J] \rightarrow \int \frac{\big[dC_1\, n_1(C_1)\big]^{\qq}}{Q!}\det{}_{\!0} (J)\,.
\end{align}
For ease of notation, we assume $Q > 0$  as anti-instantons with $Q<0$ can be treated similarly. The factor of $Q!$ arises because the constituent-instantons can be treated as identical particles. The collective coordinate measure for the $i$-th constituent-instanton is $dC_i = d\rho_i\, d^4z_i\, dU_i$. $dU_i$ is the Haar measure of the coset space $SU(N_c)/\mathcal{I}_{N_c}$, where the stability group of the instanton $\mathcal{I}_{N_c}$ is given by all $SU(N_c)$-transformations that leave the instanton unchanged. We emphasize that in the small limit the instanton density only depends upon the sizes $\rho_i$.

For small constituent-instantons, using the methods of \cite{Osborn:1978rn, Corrigan:1978ce}, the zero modes of a $Q$-instanton are simply given by the corresponding zero modes for $Q =1$, and so the quark zero mode determinant factorizes, $Z^{\qq}[J] = (Z^{(1)}[J])^Q/Q! $. Thus, for a dilute gas of $Q$-instantons, the effective Lagrangian which results is the
$Q^\text{th}$ power of the 't Hooft determinant, where {\it each} determinant is integrated over space-time,
$\sim\big[\int\!d^4x \det\big(\bar\psi_L(x) \psi_R(x)\big)\big]^Q$.

What we require, however, is a local interaction, given by a {\it single}
integral over space-time for the $Q^\text{th}$ power of the 't Hooft determinant,
$\sim\int\!d^4x\big[\det\big(\bar\psi_L(x) \psi_R(x)\big)\big]^Q$.
To find this, one needs to account for the overlap between the constituent-instantons. 
To order $\rho^4/(R^2)^2$ the only change we need to account for is the difference in the quark zero modes \cite{Corrigan:1977ma,*Christ:1978jy,*Bernard:1978ea}.
The zero mode for the $Q = 1$ instanton is
\begin{align}\label{eq:1zero}
  \psi_{fi}(x,z_i) = \frac{\sqrt{2}}{\pi}\frac{U_i \rho_i}{[(x-z_i)^2+\rho_i^2]^{3/2}}
  \frac{\gamma_\mu (x-z_i)_\mu}{|x-z_i|}\, \varphi_R\,,
\end{align}
where $\varphi_R$ is a right-handed spinor so that the zero mode is left-handed.
It will be useful later to note that far from the instanton the quark zero mode is proportional to the
free quark propagator $\Delta(x) = \gamma_\mu x_\mu/ 2\pi^2 (x^2)^2$.

For simplicity we consider instantons with charge two.
Using the methods of 
Ref.\ \cite{Grossman:1978} to derive the quark zero modes from the ADHM-solution in the limit of small constituent-instantons,
the $2 N_f$ zero modes for $Q=2$
can be expressed in terms of the $Q=1$ zero modes as:
\begin{align}\label{eq:nicezeromodes}
\begin{split}
\psi_{f1}^{(2)} = \psi_{f1} - \mathbb{X}_1 \widehat{\psi}_{f2}\,, \qquad
\psi_{f2}^{(2)} = \psi_{f2} - \mathbb{X}_2 \widehat{\psi}_{f1} \,,
\end{split}
\end{align}
where $\widehat{\psi}_{fi} = U_j U_i^\dagger \psi_{fi}$ and
\begin{align}\label{eq:overlap}
\mathbb{X}_i(x,z_i) = \frac{\rho_1\rho_2\, |x-z_i|}{[(x-z_i)^2+\rho_i^2]^{3/2}}\,.
\end{align}
So for small constituent-instantons the $Q = 2$ zero modes decompose into separate $Q = 1$ zero modes, connected by the overlap term $\mathbb{X}_i$. 

In general, the determinant depends on the locations of the constituent-instantons, $z_1$ and $z_2$, which can be rewritten as an average position $z = (z_1 + z_2)/2$ and their separation, $R_{12}$.
Integrating over $R_{12}$ the zero mode determinant becomes
\begin{align}\label{eq:detres}
\begin{split}
&\det{}_0(J)\propto\\
&\quad \mathcal{I}_{N_f} \prod_{i=1}^{2} \int_{U_i} \prod_{f = 1}^{N_f} 
\int_{x_{fi}} \psi_{fi}^\dagger(x_{fi},z) J(x_{fi}) \psi_{fi}(x_{fi},z)\,,
\end{split}
\end{align}
where $\mathcal{I}_{N_f}$ measures the overlap of the zero modes,
\begin{align}
\mathcal{I}_{N_f} = \int\! d^4R_{12}\prod_f  \mathbb{X}_i^2(x_{fi},R_{12})\,.
\end{align}
For one flavor the overlap integral is infrared-divergent, requiring a cutoff for large distances $|x-z_i|$.
Presumably, this cutoff is set by the average separation between an instanton and an anti-instanton.
For two or more flavors, a local interaction is generated when all quark zero modes are close
to the same constituent-instanton 
\footnote{To be precise, this configuration is given by
  $\rho_i \ll |x_{fi}-z_1| \ll |x_{fi}-z_2| \approx |R_{12}|$.
  For other source locations $x_{fi}$, corrections to the non-local interaction are generated,
  see Eq. (\ref{eq:OINf})}, and we find:
\begin{equation}
  \mathcal{I}_{N_f \geq 2} = \pi^2 \, \frac{(N_f+1)! (2 N_f - 3)!}{(3N_f-1)!}\, 
  \rho_1^{2 N_f} \rho_2^{4 - 2N_f} \; .
\end{equation}
Because zero modes approach free quark propagators at large distances, Eq. \eq{eq:1zero},
the zero mode determinant in Eq. \eq{eq:detres} has the form of a $2 N_f Q$-quark correlation function.
Hence, in direct generalization of \cite{tHooft:1976rip, tHooft:1976snw, tHooft:1986ooh}, the generating functional in the presence of small constituent-instantons gives rise to an effective interaction between $4 N_f$ quarks.
Assuming that the topological fluctuations are described by a dilute gas of instantons,
the contribution from dilute $Q=2$ instantons and anti-instantons generates an anomalous contribution to the
local effective Lagrangian, as a product of operators which are color singlet
\footnote{There are further interactions which while color singlet overall, are products of operators that are not;
  see the discussion before Eq. (\ref{eq:uintnf2}).  These are relevant for color superconductivity, and are ignored here.}:
\begin{align}\label{eq:effac2}
\begin{split}
\Delta\mathcal{L}_\text{eff}^{(2)} =
\frac{-\kappa_2}{K_{2,N_f}}\Big(\big[ \det_{fg}(\bar\psi_f \mathbb{P}_R \psi_g)\big]^2
+\big[ \det_{fg}(\bar\psi_f \mathbb{P}_L \psi_g)\big]^2\Big)\,,
\end{split}
\end{align}
where $\mathbb{P}_{R/L} = (1\pm\gamma_5)/2$ are the right-/left-handed projection operators and $K_{Q,N_f} = (Q!)^{N_f}/(Q N_f)!$ is a combinatorial factor.
The effective coupling in this semi-classical analysis is,
\begin{align}\label{eq:kappa2}
\kappa_2 &= (8\pi^2)^{2 N_f}\!
\int_{\rho_1}\!\! n_1(\rho_1) \rho_1^{5 N_f}\!
\int_{\rho_2}\!\! n_1(\rho_2) \rho_2^{N_f}\, \mathcal{I}_{N_f}\,.
\end{align}
This result generalizes the instanton-induced local interaction to topological charge $Q = 2$. We note that, while the effective action induced by a single instanton breaks $U(1)_A$ down to the cyclic group $\mathbb{Z}_{N_f}$, the $Q=2$ contribution has a larger residual $\mathbb{Z}_{2 N_f}$-symmetry. The computation outlined here can be generalized to arbitrary topological charge.
\\

\textbf{A low energy model.}
To illustrate the physical effect of interactions induced by higher topological charge, we consider a linear sigma model for $N_f = 2$ that includes all anomalous interactions up to quartic order. These are generated in a dilute gas of instantons and anti-instantons with $Q=1$ and 2. Classically, the global chiral symmetry of $\mathcal{G}_\text{cl} = SU(2)_L \times SU(2)_R \times U(1)_A$ is reduced to $\mathcal{G}_\text{A} = SU(2)_L \times SU(2)_R \times \mathbb{Z}_{N_f}$ by topological fluctuations. Effective mesons are given by $\Phi = (\sigma + i \eta) + (\vec{a}_0 + i \vec{\pi}) \vec{\tau}$, with the Pauli-matrices $\vec{\tau}$.
The resulting Lagrangian is a sum of two terms \cite{Jungnickel:1996aa, *Grahl:2013pba},
\begin{align}\label{eq:DL}
\begin{split}
\mathcal{L}_{\mathcal{G}_\text{cl}} &= \tr\big(\partial_\mu \Phi^\dagger\big)\big( \partial_\mu \Phi\big) + m^2\, \Tr\, \Phi^\dagger\Phi\\
&\quad + \lambda_1\, \Tr \big(\Phi^\dagger\Phi\big)^2 + \lambda_2 \big(\Tr\, \Phi^\dagger\Phi\big)^2\,,\\[1ex]
\mathcal{L}_{\mathcal{G}_\text{A}} &= -\chi_1 \big(\det\Phi + \det\Phi^\dagger\big)\\
&\quad -\chi_2 \big[  \big(\det\Phi\big)^2 + \big(\det\Phi^\dagger\big)^2 \big]\,.
\end{split}
\end{align}
We emphasize that taking into account the contributions from instantons and anti-instantons is necessary to ensure $\mathcal{CP}$-invariance.  
The term $\sim\chi_1$ arises from bosonizing the usual 't Hooft determinant from instantons with $Q = \pm 1$, while 
the term $\sim\chi_2$ is generated by bosonizing interactions with $Q = \pm 2$ in \Eq{eq:effac2}
\footnote{There is one additional anomalous quartic term, $\big(\Tr\, \Phi^\dagger\Phi\big)\big(\det\Phi + \det\Phi^\dagger\big)$, which is induced by instantons with $Q = 1$. A detailed analysis shows that, while it changes the numerical values of the couplings, the behavior of the masses as a function of the reduced temperature does not change. For details, see \App{app:low}.}. 

\begin{figure}[t]
  \centerline{  \includegraphics[width=\columnwidth]{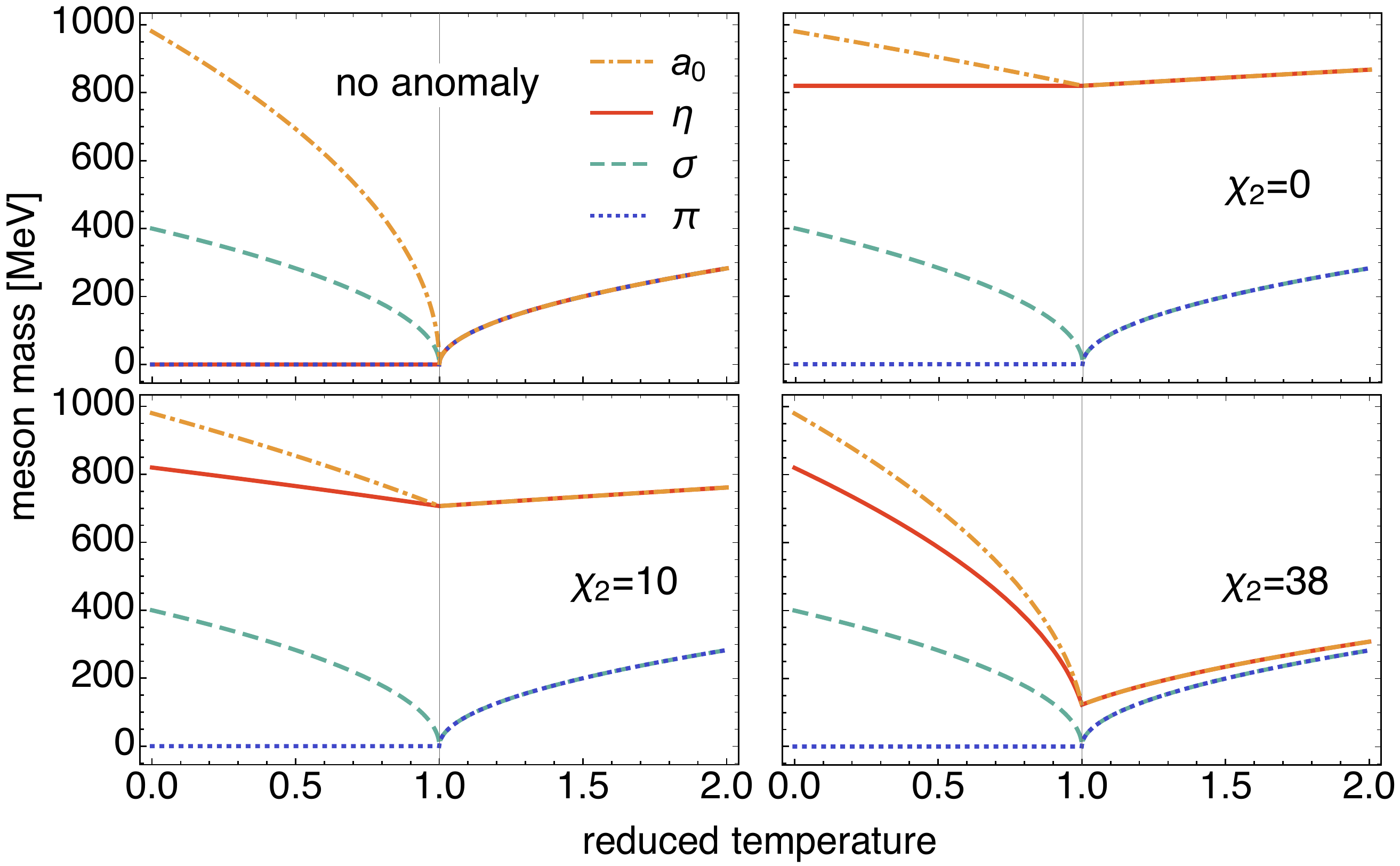} }
  \caption{The masses of mesons as a function of the reduced temperature, \Eq{eq:tred}, for two massless quarks in
    the mean-field approximation. The top left plot shows the spectrum for a $U(1)_A$-symmetric theory. The top right plot is the conventional case where the axial anomaly is induced by instantons of topological charge one. The two figures on the bottom show the effect of additional anomalous symmetry breaking due to instantons of topological charge two for two different values of the corresponding coupling $\chi_2$. Unless $\chi_2$ is very large, this is negligible.}
  \label{fig:masses}
\end{figure}

We focus on the mass spectrum of mesons in the mean-field approximation. We use the $\sigma$-, $\eta$-, $a_0$-meson masses and $f_\pi$ to fix four of the five parameters of $\mathcal{L}$ in the vacuum. Chiral symmetry breaking is controlled by the mass parameter $m^2$. By varying $m^2$ relative to its vacuum value, in \Eq{eq:tred} we define a reduced temperature $t = t(m^2)$, where $t = 0$ corresponds to the vacuum and $t = 1$ to the chiral phase transition. 
By choosing $\chi_2$ as a free parameter, we can study the impact of the topological charge-two term on the masses in the phases with broken and restored chiral symmetry.
The resulting mass spectrum is shown in \Fig{fig:masses}. The details of the computation can be found in \App{app:low}.

The splitting between the pion and eta mass is due exclusively to the axial anomaly in the chiral limit. Since $\chi_2$ is a quartic coupling, its contribution to the masses is proportional to the chiral condensate. As the condensate melts, this contribution vanishes so that $\chi_1$ is the only anomalous contribution to the masses in the symmetric phase. The larger we choose $\chi_2$,
the smaller $\chi_1$ has to be to reproduce the correct vacuum masses.
In the chirally symmetric phase $m_\sigma = m_\pi$ and $m_\eta = m_{a_0}$, but $m_\sigma \neq m_\eta$ when $\chi_1 \neq 0$.
Even when $\chi_1$ is small, however, we stress that there are still anomalous effects in the chirally symmetric phase from nonzero $\chi_2$. These manifest themselves in correlation functions of quartic and higher order.

Needless to say, the effects generated by anomalous coupling from instanton with $Q = \pm 2$ depend upon how large it is in vacuum and how rapidly it decreases with temperature $T$ and quark chemical potential $\mu$. In vacuum, the nature of the dominant fluctuations in topological charge is certainly a formidable problem in non-perturbative physics.
While this could be done on the lattice \cite{Aoki:2012yj,*Cossu:2013uua,*Fukaya:2015ara,*Tomiya:2016jwr,*Aoki:2017paw,Hashimoto:2008xg,*Bruckmann:2009pa,*Ilgenfritz:2012gu,*Bornyakov:2013iva,*Bornyakov:2014esa,*Bornyakov:2015xao,*DiGiacomo:2015eva,*Frison:2016vuc,*Bornyakov:2017crk,*Itou:2018wkm,*Jahn:2018dke,*Giusti:2018cmp,Bazavov:2012qja,*Buchoff:2013nra,*Dick:2015twa,*Petreczky:2016vrs,diCortona:2015ldu,*Bonati:2015vqz,*Borsanyi:2015cka,Brandt:2016daq,Vicari:2008jw,Bonati:2013tt} or with functional methods \cite{Mitter:2014wpa, *Braun:2014ata, *Cyrol:2017ewj, *Fischer:2018sdj, *Fu:2019hdw},
to estimate these effects we use a simple gas of dilute instantons. 
To this end, we adopt a crude bosonization scheme,
\begin{align}
2 M^2 \Phi = (\bar\psi \psi + \bar\psi \gamma_5 \psi) + (\bar\psi \vec{\tau} \psi + \bar\psi \gamma_5 \vec{\tau} \psi)\vec{\tau}\,,
\end{align}
which yields simple relations between the anomalous meson couplings in Eq. \eq{eq:DL}, and the corresponding quark couplings in the dilute instanton gas:
\begin{align}\label{eq:chis}
\chi_1 = \frac{\kappa_1 M^4}{2 K_{1,2}}\,, \qquad \chi_2 = \frac{\kappa_2 M^8}{4 K_{2,2}}\,,
\end{align}
with $\kappa_1 = \int_\rho n_1(\rho) (8 \pi^2)^{N_f} \rho^{3 N_f}$ \cite{tHooft:1976rip, tHooft:1976snw} and
$\kappa_2$ is given in Eq. \eq{eq:kappa2}. The mass scale $M$ is a fundamental parameter of our effective theory.
Motivated by the complete computation
at one loop order \cite{tHooft:1976rip,*tHooft:1976snw,*tHooft:1986ooh}, and the partial computation at two loop order
\cite{Morris:1984zi,*Ringwald:1999ze}, for three colors and two massless flavors we take the density of
a single instanton in the vacuum to be
\begin{align} \label{instanton_density}
n_1(\rho) = \frac{d_{\overline{MS} } }{\rho^5} \; \left( \frac{8 \pi^2}{g^2}   \right)^6 \; 
\exp\left( - \; \frac{8 \pi^2}{g^2 } \right)\,,
\end{align}
where $g^2 = g^2(\rho \Lambda_{\overline{MS}})$ is the running coupling constant at two loop order and $d_{\overline{MS}}$ is a renormalization-scheme dependent constant, $\approx 0.00449$ for $N_c = 3$ and $N_f = 2$.
The apparent simplicity of our form for the instanton density belies a major assumption 
that everywhere the coupling $g^2$ appears that we can replace it with
$g^2(\rho \Lambda_{\overline{MS} })$.

This assumption, while admittedly extreme, is both simple and useful. 
Owing to the interplay between the running coupling from the classical action in the exponential and the factor $\sim g^{-12}$ from the collective coordinate Jacobian, $n_1(\rho)$ develops a pronounced maximum
at $\rho \Lambda_{\overline{MS}} \approx 1/2$.
For typical values of $\Lambda_{\overline{MS}} \approx 300$\,MeV \cite{Tanabashi:2018oca}, this implies typical instanton sizes of $\rho \approx 1/3$\,fm, which is consistent with the value in an instanton liquid
\cite{Shuryak:1978yk,*Shuryak:1981ff,*Shuryak:1982dp,*Shuryak:1982hk,*Shuryak:1982qx,*Shuryak:1992ke,*Ilgenfritz:1988dh,*Ilgenfritz:1994nt,*Schafer:1995pz,*Rapp:1997zu,*Rapp:1999qa,*Shuryak:2000df,*Liu:2015ufa,*Liu:2015jsa,*Liu:2016mrk,*Liu:2016thw,*Liu:2016yij,*Liu:2018znq,Schafer:1996wv,Diakonov:1979nj,*Diakonov:1983hh,*Diakonov:1984vw,*Diakonov:1985eg,*Diakonov:1986tv,*Diakonov:2002fq,*Diakonov:2004jn,*Diakonov:2005qa,*Gromov:2005ij,*Diakonov:2007nv,*Bruckmann:2009nw,Diakonov:2009jq}.
Of course we cannot
compute reliably at large $\rho$, since inevitably the instanton size is comparable to the confinement scale,
and semiclassical approximations break down. 

Since the two anomalous couplings $\chi_{1}$ and $\chi_2$
only depend on a single free parameter $M$, we can redo the mean-field analysis
of the meson masses, and find a unique value for $M$ in the vacuum. From the dilute instanton gas, \Eq{instanton_density}, with
$\Lambda_{\overline{MS}} = 0.3$\,GeV,
\begin{equation}
  \kappa_1 = 3886\, \text{GeV}^{-2}\,, \quad \kappa_2 = 1.819 \times 10^8\, \text{GeV}^{-8} \, .
\end{equation}
In vacuum, this gives $M = 0.0953\, \text{GeV}$, so from \Eq{eq:chis},
\begin{equation}
\chi_1 = 0.320\, \text{GeV}^2\,, \quad \chi_2 = 1.852 \,.
\end{equation}
Of couse, $\chi_1$, $\chi_2$ and {\it all} other anomalous effects are {\it very}
sensitive to the value chosen for $\Lambda_{\overline{MS}}$.  Nevertheless, we expect that our naive
computations should give a reasonable estimate for their overall magnitude.

We conclude by discussing how the dilute instanton gas evaporates as $T$ and $\mu$ increase. For a single instanton we approximate the change to the instanton density for three colors and two flavors as
\begin{equation}
  n_1(\rho,T,\mu) = \exp\!\bigg[- \frac{2 \pi^2}{g^2}\rho^2 m_D^2 
    - 14 A(\pi \rho T) \bigg] n_1(\rho)\, ,
  \label{density_Tmu1}
\end{equation}
where $ m^2_D(T,\mu)$ is the Debye mass at leading order, Eq. \ref{debye_mass}, and
$A(x)$ is given in Eq. \ref{ax} \cite{Gross:1980br,Altes:2014bwa,*Altes:2015wla}. Owing to the screening of the color-electric field in the medium, the instanton density decreases both with increasing $T$ and $\mu$. We find that instanton effects are decreased to 10\% of their strength in vacuum at about $T \approx 0.7\, \Lambda_{\overline{MS}}$ at $\mu = 0$ and $\mu \approx 2.4\, \Lambda_{\overline{MS}}$ at $T = 0$. Using realistic values for the critical temperature $T_c$ \cite{Aoki:2006we,*Aoki:2006br,*Aoki:2009sc,*Bazavov:2011nk,*Borsanyi:2012ve,*Bazavov:2014pvz,*Bhattacharya:2014ara} and $\Lambda_{\overline{MS}}$ \cite{Tanabashi:2018oca}, we find that instanton effects are significantly suppressed at temperatures $T \gtrsim 1.5 T_c$ for $\mu = 0$, consistent with lattice results \cite{Aoki:2012yj,*Cossu:2013uua,*Fukaya:2015ara,*Tomiya:2016jwr,*Aoki:2017paw,Hashimoto:2008xg,*Bruckmann:2009pa,*Ilgenfritz:2012gu,*Bornyakov:2013iva,*Bornyakov:2014esa,*Bornyakov:2015xao,*DiGiacomo:2015eva,*Frison:2016vuc,*Bornyakov:2017crk,*Itou:2018wkm,*Jahn:2018dke,*Giusti:2018cmp,Bazavov:2012qja,*Buchoff:2013nra,*Dick:2015twa,*Petreczky:2016vrs,diCortona:2015ldu,*Bonati:2015vqz,*Borsanyi:2015cka,Brandt:2016daq,Vicari:2008jw,Bonati:2013tt}.
As discussed in \App{app:diga}, at zero temperature in a dilute instanton gas, instantons evaporate only at \emph{extremely} high densities of $\mu \gtrsim 1.5\, \pi T_c$. Using $T_c = 156$\,MeV, this corresponds to baryon chemical potentials of
$\mu_B \gtrsim 2$\, GeV. 
\\

\textbf{Summary \& outlook.}
We demonstrated that novel effective interactions are generated by instantons of higher topological charge. In general, instantons of topological charge $Q$ give rise to $2 N_f |Q|$-quark interactions. This opens up the possibility to study the effects of the axial anomaly \emph{directly} for higher correlation function of quarks or hadrons. Besides the example studied here, 
it is especially interesting to study QCD with one light flavor,
where instantons with $Q = \pm 2 $ generate a mass for the $\eta$ meson.
These methods can also be used to compute anomalous couplings for heterochiral mesons with $ J \geq 1$ \cite{Giacosa:2017pos} and
tetraquark mesons \cite{Pisarski:2016ukx}.
\\

\textit{Acknowledgements.}
We thank S.\ Mukherjee for originally asking us about
interactions induced by instantons.  We also
thank F.\ Karsch, R.\ Larsen, P.\ Petreczky, and E.\ Shuryak for discussions.
R.D.P. is funded by the U.S. Department of Energy under contract DE-SC0012704;
F.R. is supported by the Deutsche Forschungsgemeinschaft (DFG) through grant RE 4174/1-1
and in part by the U.S. Department of Energy under contract DE-SC0012704.

\newpage
\begin{widetext}
\appendix

\vspace{5ex}
\centerline{\textbf{\large Appendix}}

\section{Conventions}\label{app:conv}

We use a chiral representation for the Euclidean gamma matrices: with the Pauli matrices $\sigma^i$
\begin{align}\label{eq:simu}
\sigma^\mu = (-i \mathds{1}_2, \vec{\sigma})^\mu\,,\quad \bar\sigma^\mu = (i \mathds{1}_2, \vec{\sigma})^\mu\,,
\end{align}
then
\begin{align}
\gamma^\mu =
\begin{pmatrix}  
0 & i \sigma^\mu \\
-i \bar\sigma^\mu & 0
\end{pmatrix}\, 
\end{align}
and
\begin{align}
\gamma^5 = \gamma^0 \gamma^1 \gamma^2 \gamma^3 =
\begin{pmatrix}  
-\mathds{1}_2 & 0 \\
0 & \mathds{1}_2
\end{pmatrix}\,.
\end{align}
Left- (right-) handed fields have eigenvalue $-1$ ($+1$) with respect to $\gamma^5$.
The projection operators on left- and right-handed fields are given by
\begin{align}\label{eq:chiproj}
\mathbb{P}_{L/R} = \frac{\mathds{1}_4 \mp \gamma^5}{2}\,.
\end{align}
The matrices in \Eq{eq:simu} can be used to define the basis quaternions
\begin{align}\label{eq:basquat}
\bar \alpha^\mu = i \sigma^\mu\,, \qquad \alpha^\mu = -i \bar\sigma^\mu\,.
\end{align}
We also define
\begin{align}\label{eq:sigmamunu}
\begin{split}
\sigma^{\mu\nu} = \frac{1}{2} \big( \sigma^\mu \bar\sigma^\nu - \sigma^\nu\bar\sigma^\mu \big)\,,
\qquad
\bar\sigma^{\mu\nu} = \frac{1}{2} \big(\bar\sigma^\mu \sigma^\nu - \bar\sigma^\nu \sigma^\mu \big)\,,
\end{split}
\end{align}
which are selfdual and anti-selfdual respectively,
\begin{align}
\begin{split}
\sigma^{\mu\nu} = + \frac{1}{2} \epsilon^{\mu\nu\rho\sigma} \sigma^{\rho\sigma}\,, 
\qquad
\bar\sigma^{\mu\nu} = - \frac{1}{2} \epsilon^{\mu\nu\rho\sigma} \bar\sigma^{\rho\sigma}\,.
\end{split}
\end{align}
They are related to the 't Hooft symbols $\eta^{a\mu\nu}$ through the $SU(2)$ color generators $T^a$
\begin{align}
\begin{split}
\sigma^{\mu\nu} = -2 \eta^{a\mu\nu} T^a\,,
\qquad
\bar\sigma^{\mu\nu} = -2 \bar\eta^{a\mu\nu} T^a\,.
\end{split}
\end{align}
We distinguish between Pauli matrices $\sigma^i$
for the Dirac matrices, and the $\tau^a$ in color space, with $T^a = - i \tau^a/2$.
For $N_c > 2$ these generators are given by embedding $SU(2)$ into $SU(N_c)$.
The 't Hooft symbols are
\begin{align}
\begin{split}
\eta^{a\mu\nu} = \delta^{a\mu} \delta^{\nu 4} - \delta^{a\nu} \delta^{\mu 4} + \epsilon^{a\mu\nu}\,,
\qquad
\bar\eta^{a\mu\nu} = \delta^{a\nu} \delta^{\mu 4} - \delta^{a\mu} \delta^{\nu 4} + \epsilon^{a\mu\nu}\, ,
\end{split}
\end{align}
and are also (anti-) selfdual.

\section{Small instantons from the ADHM-construction}

The most general form of an instanton with charge $\pm Q$
is obtained by means of the ADHM construction \cite{Atiyah:1978ri, Christ:1978jy}.
The $Q$-instanton solution is described by a superposition of instantons with unit charge,
where each of the constituent-instantons is characterized by a position $z_i$,
a size $\rho_i$, and its orientation in the gauge group, parametrized by a matrix $U_i$.
We consider the limit in which the distance between the constituent-instantons is large relative
to their scale sizes, $|z_i - z_j| \gg \rho_i$ for all $i \neq j$, which we term {\it small}.
For a systematic expansion of the ADHM solution for small constituent-instantons see Ref.\ \cite{Christ:1978jy}. Since this is relevant for the construction of the quark zero modes, we outline the construction here.

Without loss of generality, we assume $Q > 0$. Anti-instantons can always be obtained trivially be replacing the selfdual matrices $\sigma^{\mu\nu}$, $\eta^{a\mu\nu}$ by the corresponding anti-selfdual matrices $\bar\sigma^{\mu\nu}$, $\bar\eta^{a\mu\nu}$, see \App{app:conv}. 
The most general self-dual gluon field with topological charge $Q$ can be constructed algebraically from a $(Q+1)\times Q$ matrix $M(x)$ whose entries are quaternionic. Each matrix element $M_{ab}$ can therefore be viewed as a $2\times 2$ matrix,
\begin{align}\label{eq:mquat}
M_{ab} = \alpha_\mu M^\mu_{ab}\,,
\end{align}
where the basis quaternions $\alpha$ are defined in \Eq{eq:basquat}. $M(x)$ is chosen to be linear in spacetime $x$,
\begin{align}\label{eq:mans}
M(x) = B - C x\,,
\end{align}
with the constant quaternionic $(Q+1)\times Q$ matrices $B$, $C$ and the quaternionic spacetime coordinate $x = \alpha_\mu x_\mu$. $M$ is required to obey the reality condition
\begin{align}\label{eq:mreal}
M^\dagger(x) M(x) = R(x)\,,
\end{align}
where $R$ is a real $Q\times Q$ quaternionic matrix. Hence, each entry is proportional to $\alpha_0 = \mathds{1}_2$. The quaternionic conjugate ${}^\dagger$ is given by,
\begin{align}
\big(M^\dagger\big)_{ab}^0 = M_{ba}^0\,,\qquad \big(M^\dagger\big)_{ab}^i = - M_{ba}^i\,.
\end{align}
When expressed in terms of \Eq{eq:mquat}, $M$ can be viewed as a complex $2(Q+1)\times 2Q$ matrix. We choose the quaternionic representation for convenience.
Aside from $M$, the other crucial ingredient is the quaternionic $(Q+1)$ column vector $N(x)$, which obeys
\begin{align}\label{eq:neq}
N^\dagger(x) M(x) = 0\,, \qquad N^\dagger(x) N(x)= \mathds{1}_2\,.
\end{align}
The first equation yields $Q$ equations for the $(Q+1)$ entries of $N$, so $Q$ entries of $N$ can always be expressed in terms of one other entry. The choice of this entry corresponds to a gauge choice for the $Q$-instanton. The second equation is a normalization condition.
By solving Eqs.\ \eq{eq:mreal} and \eq{eq:neq} with the ansatz \eq{eq:mans}, the $SU(2)$ $Q$-instanton is given by
\begin{align}\label{eq:adhminst}
A_\mu^{(Q)}(x) = N^\dagger(x) \partial_\mu N(x)\,,
\end{align}
where $N$ is determined up to $8Q-3$ free parameters, corresponding to the complete set of collective coordinates of the instanton for $N_c = 2$. Only the relative orientation of the constituent-instantons in the gauge group is counted, leaving three parameters for the overall gauge rotation of the solution, so $8 Q$ parameters in all. We follow the explicit construction of $M$ and $N$ in Ref.\ \cite{Christ:1978jy}. The first column of $M$ is given by a vector of $Q$ constant quaternions $q = (q_1,\dots, q_Q)$,
\begin{align}
M_{1j} = q_j\,,\quad j\in[1,\dots, Q]\,,
\end{align}
and the remaining $Q\times Q$ elements of $M$ are given by
\begin{align}\label{eq:mhat}
\widehat{M}_{ij}(x) = \delta_{ij} (z_i - x ) + b_{ij}(x)\,.
\end{align}
$b_{ij}$ is a quaternionic matrix. The diagonal elements of $\widehat{M}$ are parametrized by $z_i$, so $b_{ii} = 0$. We show below that $z_i$ can be interpreted as instanton locations.
The reality constraint in \Eq{eq:mreal} is fulfilled if $b_{ij}$ is symmetric, $b_{ij} = b_{ji}$, and obeys
\begin{align}\label{eq:bijeq}
\frac{1}{2} ( q_i^* q_j - q_j^* q_i ) + (z_i - z_j)^* b_{ij} + \frac{1}{2} \sum_{k=1}^Q (b_{ki}^* b_{kj} - b_{kj}^* b_{ki}) = R_{ij}\,.
\end{align}
The column vector $N$ is
\begin{align}
N(x) = \frac{1}{\sqrt{\xi}}
\begin{pmatrix}
u \\
- \Big[ \big( \widehat{M}^\dagger\big)^{-1} q^\dagger \Big]_1 \cdot u \\
\vdots \\
- \Big[ \big( \widehat{M}^\dagger\big)^{-1} q^\dagger \Big]_Q \cdot u
\end{pmatrix}\,,
\end{align}
where $u$ is an arbitrary, possibly $x$-dependent, unit quaternion. Different $u$ correspond to gauge-equivalent solutions, where $u = \alpha_0$ corresponds to singular gauge. $\xi$ is determined from the normalization condition in \Eq{eq:neq},
\begin{align}\label{eq:xi}
\xi = 1 + q \widehat{M}^{-1}\big(\widehat{M}^{\,\dagger}\big)^{-1} q^\dagger\,.
\end{align}
It is left to specify the $b_{ij}$, {\it i.e.}\ to solve \Eq{eq:bijeq}. For this, we restrict ourselves to the limit of small constituent-instantons as this is all we need for our purposes. We first note that every quaternion $q_i$ can be represented by a modulus and a phase,
\begin{align}\label{eq:quatscale}
q_i = |q_i| U_i = \sqrt{q_i^* q_i}\, U_i = \rho_i\, U_i\,,
\end{align}
where $U_i$ is an $SU(2)$ matrix, and there is no summation over $i$ here. The magnitude of $q_i$ can be interpreted as the scale of the instanton, $\rho_i = |q_i|$. $U_i$ parametrizes the orientation in the gauge group. In the limit of small constituent-instantons, we introduce a small parameter $\zeta$ and replace
\begin{align}
q_i \rightarrow \zeta\, q_i\,,
\end{align}
where $q_i$ is kept fixed. We then consider the instanton scale to be small relative to the instanton separations, $\zeta |q_i| \ll |z_i - z_j|$, and expand \Eq{eq:bijeq} to leading order in $\zeta$. The solution is unchanged if
$\widehat{M} \rightarrow T \widehat{M}$, where $T$ is an orthogonal quaternionic matrix. Choosing
\begin{align}
T_{ij} = \delta_{ij} + \frac{R_{ij}}{(z_i - z_j)^2} + \mathcal{O}(\zeta^4)\,,
\end{align}
cancels the right-hand side of \Eq{eq:bijeq}. To this order, then, one finds
\begin{align}
b_{ij} = \frac{1}{2} \frac{z_i - z_j}{(z_i - z_j)^2} ( q_i^* q_j - q_j^* q_i )\,.
\end{align}
Thus, for small constituent-instantons we can neglect $b_{ij}$ and $M$ becomes
\begin{align}\label{eq:msmall}
M(x) = 
\begin{pmatrix}
q_1       & \cdots & q_Q \\
(z_1 -x) & \cdots & 0\\
\vdots           & \ddots & \vdots\\
0           & \cdots & (z_Q -x)
\end{pmatrix} + \mathcal{O}(\zeta^2)\,.
\end{align}
The dominant contribution to the determinant of zero modes for the $Q$-instanton comes from distances large relative to the size of each constituent-instanton $|x-z_i| \gg \rho_i$.
The constant quaternionic matrices $B$ and $C$ in \Eq{eq:mans} are then
\begin{align}\label{eq:BCexpl}
B = 
\begin{pmatrix}
q_1       & \cdots & q_Q \\
z_1 & \cdots & 0\\
\vdots           & \ddots & \vdots\\
0           & \cdots & z_Q
\end{pmatrix}\,,
\qquad
C = 
\begin{pmatrix}
0       & \cdots & 0 \\
\mathds{1}_2 & \cdots & 0\\
\vdots           & \ddots & \vdots\\
0           & \cdots & \mathds{1}_2
\end{pmatrix}\,.
\end{align} 
With this, $\xi = \xi_0 + \mathcal{O}(\xi^2)$ in \Eq{eq:xi}, where
\begin{align}\label{eq:xi0exp}
\xi_0(x) =1+ \sum_{i= 1}^{Q} \frac{\rho_i^2}{(x - z_i)^2}\,, 
\end{align}
and $N(x)$ becomes
\begin{align}\label{eq:nsmall}
N(x) = \frac{1}{\sqrt{\xi_0}}
\begin{pmatrix}
u \\
\frac{x-z_1}{(x-z_1)^2}\, q_1^*\cdot u \\
\vdots \\
\frac{x-z_Q}{(x-z_Q)^2}\, q_Q^*\cdot u
\end{pmatrix} + \mathcal{O}(\zeta^2)\,.
\end{align}
Choosing $u = \alpha_0$ and plugging this into \Eq{eq:adhminst} then yields the small $Q$-instanton,
\begin{align}\label{eq:qismall}
A_\mu^{\qq}(x) = \frac{1}{\xi_0(x) } \sum_{i=1}^{Q} U_i \bar\sigma^{\mu\nu} U_i^\dagger\, \rho_i^2\, \frac{ (x-z_i)_\nu}{|x-z_i|^4}\,,
\end{align}
with $\bar\sigma^{\mu\nu}$ defined in Eq. \eq{eq:sigmamunu}. 
If all constituent-instantons are aligned in color space, $U_i = U_j$ for all $i,j = 1,\dots,Q$, this reduces to 't Hooft's solution \cite{tHooftunpub}.

A key feature of the limit of small constituent-instantons is that the field of the $Q$-instanton, Eq. \eq{eq:qismall},
is only significant when $x$ close to the location of one of the constituent-instantons, at $z_i$.
In the vicinity of each $z_i$, Eq. \eq{eq:qismall} looks like a $Q=1$ instanton in singular gauge,
\begin{align}
A_\mu^{\qq}(x)\Big|_{|x-z_i| \approx \mathcal{O}(\rho_i)} = U_i \bar\sigma^{\mu\nu} U_i^\dagger\, \frac{\rho_i^2}{(x-z_i)^2} \frac{(x-z_i)_\nu}{(x-z_i)^2 + \rho_i^2}\,.
\end{align}
To leading order in powers of $\rho_i^2/(z_i - z_j)^2$, in the saddle-point approximation
the generating functional for a small $Q$-instanton
factorizes into a product of generating functionals in the backgrounds of single instantons, \Eq{eq:ZQdilute}.
See Refs. \cite{Osborn:1981yf, Dorey:2002ik} for a more detailed discussion of this factorization.

\section{Quark zero modes for a small $Q$-instanton}

In the presence of a $Q$-instanton quarks have zero modes,
\begin{align}
\slashed{D}^{\qq} \psi_{fi}^{\qq} = 0\,,
\end{align}
where $\slashed{D}^{\qq} = \gamma_\mu \big(\partial_\mu + A_\mu^{\qq}\big)$ is the Dirac operator in the fundamental representation; remember that $f$ is an index for flavor. The Atiyah-Singer index theorem demonstrates
that gauge field configurations with topological charge $Q$ produce
$N_f |Q|$ left-handed (for $Q>0$) or right-handed (for $Q<0$) quark zero modes \cite{Atiyah:1963zz, Coleman:1978ae}.
With the ADHM construction the zero modes are \cite{Corrigan:1978ce, Osborn:1978rn}
\begin{align}
\psi_{fi}^{\qq} =  \nu\, \big(N^\dagger C R^{-1} \big)_i\cdot \varphi\,;
\end{align}
this is a left-handed Weyl spinor, $\varphi = \epsilon/\sqrt{2}$, and $\nu$ is a normalization constant. For small
constituent-instantons with $Q = 2$ we use Eqs.\ \eq{eq:msmall}, \eq{eq:BCexpl} and \eq{eq:nsmall} to find:
\begin{align}\label{eq:CRexpl}
C = 
\begin{pmatrix}
0  & 0 \\
\mathds{1}_2 & 0 \\
0 & \mathds{1}_2
\end{pmatrix}\,,
\qquad
R =
\begin{pmatrix}
(x-z_1)^2 + \rho_1^2 & q_1^* q_2 \\
q_2^* q_1 & (x-z_2)^2 + \rho_2^2
\end{pmatrix}\,.
\end{align}
Note that to leading order $R$ is diagonal, $\text{diag}\big[(x-z_1)^2,(x-z_2)^2\big]$; the other terms are $\mathcal{O}(\zeta^2)$. Since $R$ is real, we use
\begin{align}
R^{-1} = \frac{1}{\det R} \begin{pmatrix} R_{22} & -R_{12} \\ - R_{21} & R_{11} \end{pmatrix}\,,
\end{align}
with
\begin{align}
\det R = (x-z_1)^2(x-z_2)^2+ \rho_2^2(x-z_1)^2+\rho_1^2(x-z_2)^2\,.
\end{align}
This yields
\begin{align}\label{eq:zerofull}
\begin{split}
\psi_{f1}^{(2)}(x) &= \nu\, \frac{U_1 \rho_1}{|x-z_1|} \gamma_\mu \bigg[ (x-z_1)_\mu \Big((x-z_2)^2+\rho_2^2\Big) - \rho_2^2 (x-z_2)_\mu \frac{(x-z_1)^2}{(x-z_2)^2} \bigg]\\[1ex]
&\quad\times \frac{|x-z_2|}{\big[(x-z_1)^2(x-z_2)^2+ \rho_2^2(x-z_1)^2+\rho_1^2(x-z_2)^2\big]^{3/2}}\, \varphi_R\,,\\[2ex]
\psi_{f2}^{(2)}(x) &= \psi_{f1}^{(2)}(x)\Big|_{1\leftrightarrow 2}\,,
\end{split}
\end{align}
with the right-handed spinor
\begin{align}\label{eq:spinorR}
\varphi_R^{\alpha c} = \frac{1}{\sqrt{2}}\begin{pmatrix} 0 \\ \epsilon \end{pmatrix}_{\alpha c}\,,
\end{align}
where $\alpha$ is a spinor index, $c$ is a $SU(2)$ color index and $\epsilon$ is the antisymmetric tensor.
Because of the factor of $\gamma_\mu$ in \Eq{eq:zerofull},
the zero mode $\psi_{fi}^{\qq}(x)$ is left-handed when $Q > 0$, as required by the index theorem.
For an anti-instanton, $Q < 0$, one simply has to replace $\varphi_R$ by
\begin{align}\label{eq:spinorL}
\varphi_L^{\alpha c} = \frac{1}{\sqrt{2}}\begin{pmatrix} \epsilon \\ 0 \end{pmatrix}_{\alpha c}\,.
\end{align}
To turn the Weyl into Dirac spinors we use $\gamma_\mu \varphi_R = \bar \alpha_\mu \varphi$, and \Eq{eq:quatscale} to express the quaternions $q_i$ in terms of the sizes and gauge group orientations of the
constituent-instantons. As a results, the $i$-th quark zero mode carries the gauge group orientation of
the corresponding constituent-instanton.
If they had the same orientation, $U_1 = U_2$, they would be identical to the zero modes that follow from 't Hooft's solution for the aligned instanton \cite{Grossman:1978}.

For small constituent-instantons,
\begin{align}\label{eq:limit1}
|R_{12}| \equiv |z_1 - z_2| \gg \rho_1, \rho_2\,.
\end{align}
We consider the behavior of the zero modes far from the
constituent-instantons,
\begin{align}\label{eq:limit2}
|x-z_1|,\, |x-z_2| \gg \rho_1,\, \rho_2\,. 
\end{align}
Since the $Q$-instanton is an extended object, in general it generates a quark interaction which is non-local.
We wish to extract the local term, in which the sizes of the constituent-instantons can be neglected.
In this limit, \Eq{eq:zerofull} is approximately
\begin{align}\label{eq:qzero}
\begin{split}
\psi_{f1}^{(2)}(x)
&\approx \nu \frac{U_1 \rho_1}{[(x-z_1)^2+\rho_1^2]^{3/2}} \frac{\gamma_\mu (x-z_1)_\mu}{|x-z_1|}\, \varphi_R\\[1ex]
&\quad -\nu \frac{U_1 \rho_2}{[(x-z_2)^2+\rho_2^2]^{3/2}} \frac{\gamma_\mu (x-z_2)_\mu}{|x-z_2|} \frac{\rho_1 \rho_2\, |x-z_1|}{[(x-z_1)^2+\rho_1^2]^{3/2}} \, \varphi_R\\[2ex]
&= \psi_{f1}(x,z_1) - \mathbb{X}_1(x,z_1)\, \widehat{\psi}_{f2}(x,z_2)\,,
\end{split}
\end{align}
and analogously for the second zero mode $\psi_{f2}^{(2)}(x)$. In this approximation, we drop
a term $\rho_1^2\rho_2^2$ in the denominator as subleading, and then used \Eq{eq:limit2}
to reduce the expression. We define the $Q=1$ zero modes as
\begin{align}\label{eq:psifi}
\begin{split}
\psi_{fi}(x,z_i) &= \nu\, \frac{U_i \rho_i}{\big[(x-z_i)^2+\rho_i^2\big]^{3/2}} \frac{\gamma_\mu (x-z_i)_\mu}{|x-z_i|}\, \varphi_R\,,\\[2ex]
\widehat{\psi}_{fi}(x,z_i) &= U_j U_i^\dagger \psi_{fi}(x,z_i)\,,
\end{split}
\end{align}
and
\begin{align}
\begin{split}
\mathbb{X}_i(x,z_i) = \frac{\rho_1 \rho_2\, |x-z_i|}{[(x-z_i)^2+\rho_i^2]^{3/2}}\,.
\end{split}
\end{align}
The first zero mode for $Q=2$ is a sum of a zero mode concentrated at $z_1$ plus
an overlap term, which is $\mathbb{X}_i$ times the zero mode concentrated at $z_2$; that for the second zero mode
is similar, with the exchange of $1 \leftrightarrow 2$.
The overlap term is $\mathcal{O}(\zeta^3)$. To leading order, \emph{i.e.}\ $\mathcal{O}(\zeta^1)$, the zero modes reduce to the $Q=1$ zero modes, $\psi_{f1}^{(2)} = \psi_{f1} + \mathcal{O}(\zeta^3)$.

In \Fig{fig:qzero} we compare the exact quark zero mode in the limit of small constituent-instantons, \Eq{eq:zerofull}, to the approximate form in \Eq{eq:qzero}.
For this figure, we project onto the scalar part of the zero mode $\psi_{f1}^{(2)}$ via
\begin{align}\label{eq:psiplot}
\varphi_R^\dagger\, \frac{\gamma_\mu (x-z_1)_\mu}{|x-z_1|}\, \psi_{f1}^{(2)}(x)\,,
\end{align}
for the configuration $(x-z_1) \cdot (x-z_2) = |x-z_1| |x-z_2|$, so that the function depends
only upon the relative distances.
Our approximate form is very good even close to $z_1$ and $z_2$ for $\rho/|R| \lesssim 0.3$.
At leading order the normalization constant $\nu$ is determined via
\begin{align}
\int\!d^4x\, \psi_{fi}^{(2) \dagger}(x) \psi_{fi}^{(2)}(x) \approx \int\!d^4x\, \psi_{fi}^\dagger(x,z_i) \psi_{fi}(x,z_i) \equiv 1\,,
\end{align}
which gives $\nu = \sqrt{2}/\pi$.

\begin{figure}[t]
  \centerline{  \includegraphics[width=.4\textwidth]{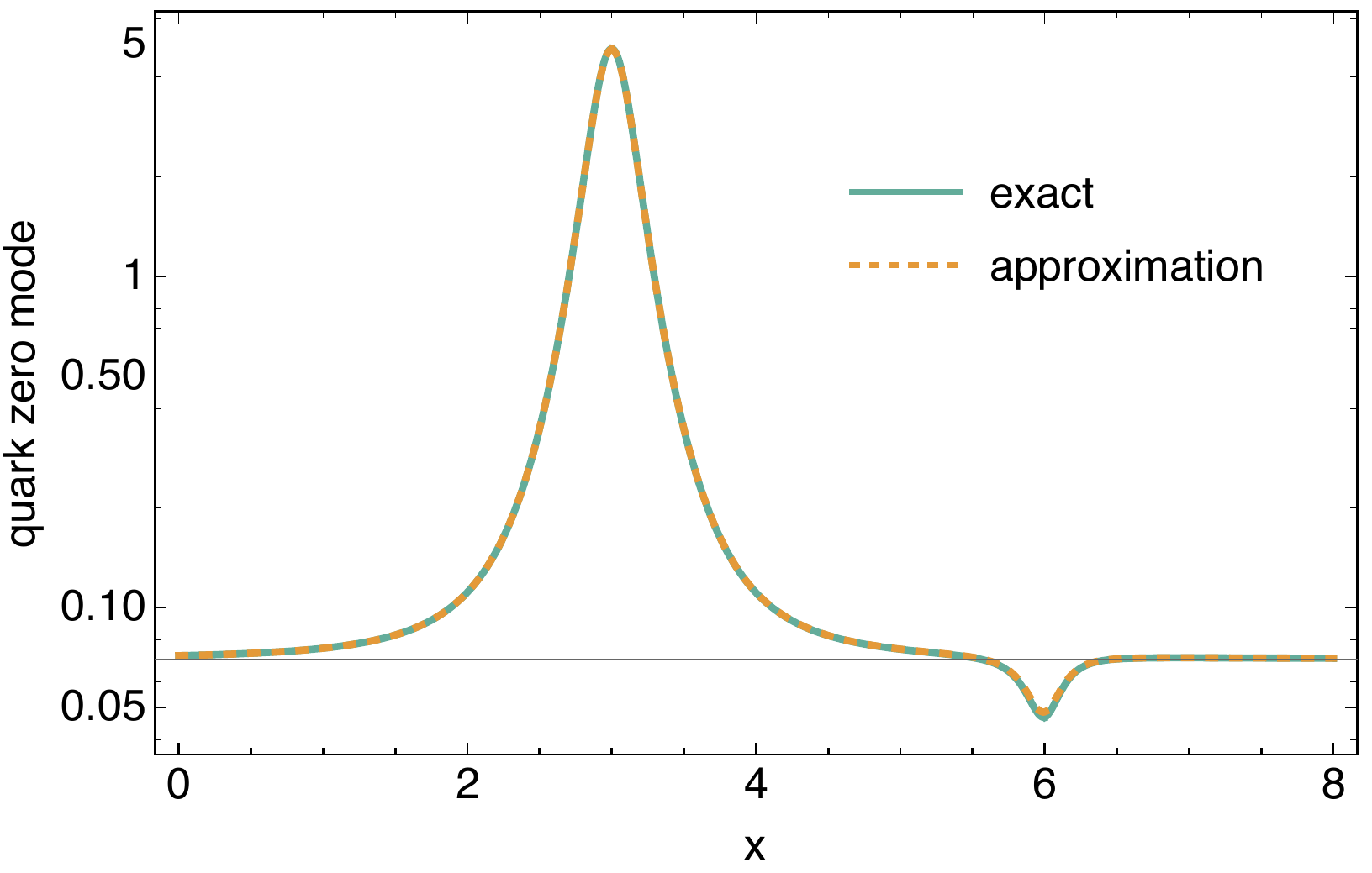} }
  \caption{Comparison between the exact form of the $Q = 2$ quark zero mode $\psi_{f1}^{(2)}(x)$ for small constituent-instantons and our approximation in
    \Eq{eq:qzero}. We used the parameters $z_1 = 3$, $z_2 = 6$ and $\rho_1 = \rho_2 = 0.21$ for this plot.
    An offset of $+0.07$ was added so that a logarithmic scale could be used along
    the $y$-axis, since the baseline of the zero mode vanishes.
    The scalar function plotted is defined in \Eq{eq:psiplot}.}
  \label{fig:qzero}
\end{figure}

To make the computations more transparent we use a graphical representation of the zero modes in \Eq{eq:qzero},
\begin{align}\label{eq:qzeropic}
\begin{split}
\psi_{f1}^{(2)}(x) &= \Big(\raisebox{-9pt}{\includegraphics[width=.07\textwidth]{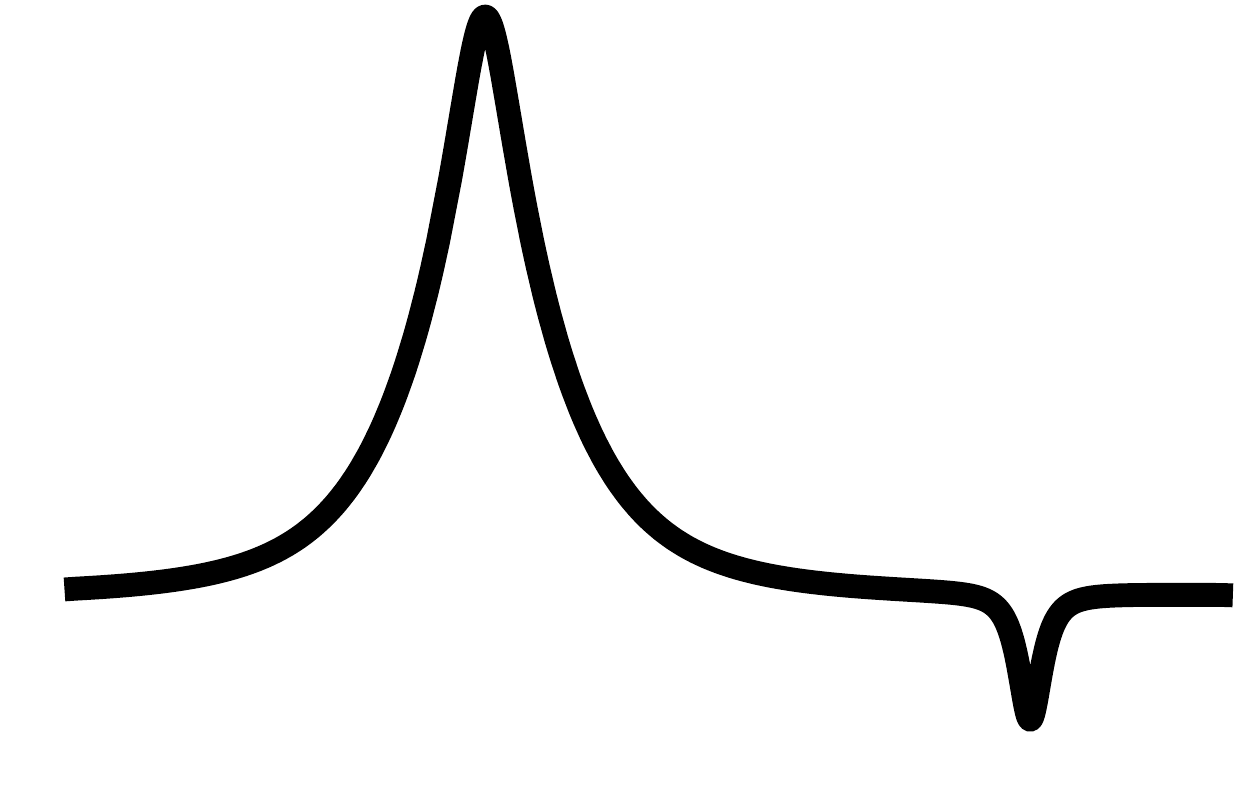}}\Big)_{f1} 
= \Big(\raisebox{-9pt}{\includegraphics[width=.07\textwidth]{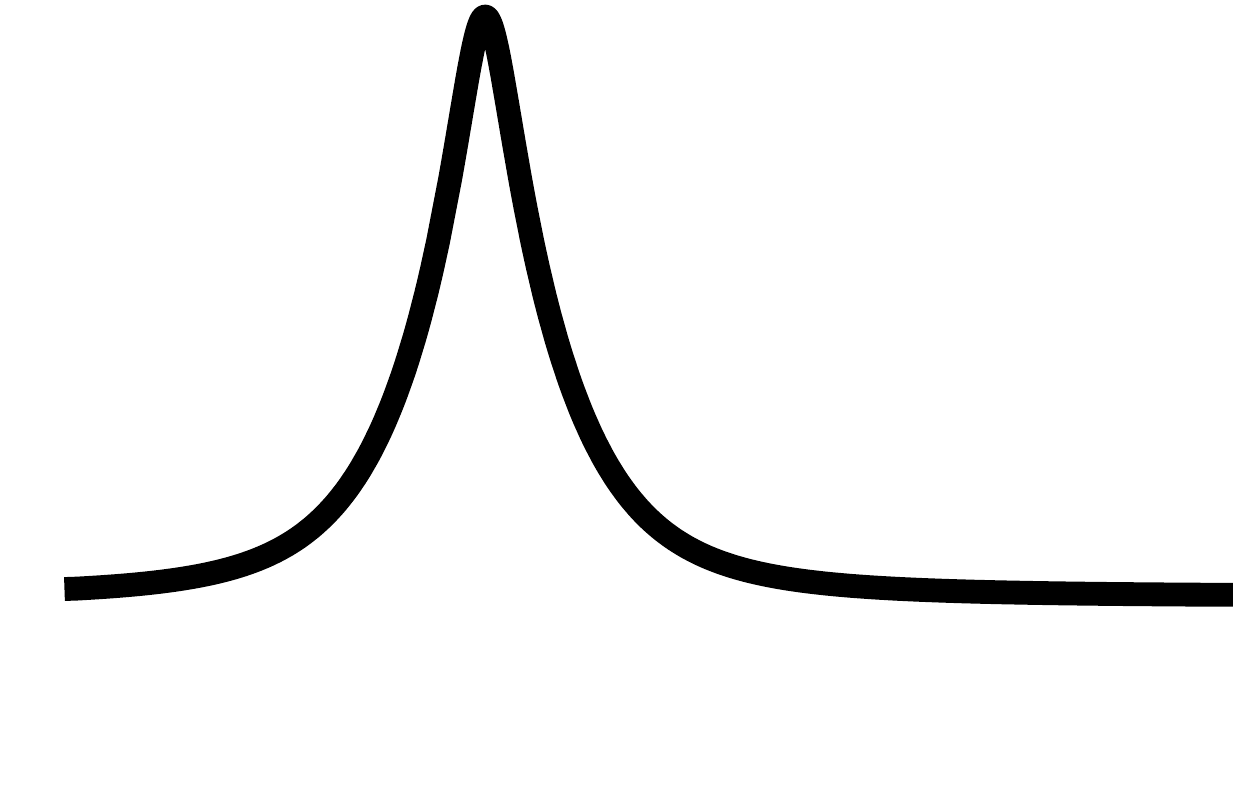}}\Big)_{f1} 
- \mathbb{X}_1 \Big(\raisebox{-9pt}{\includegraphics[width=.07\textwidth]{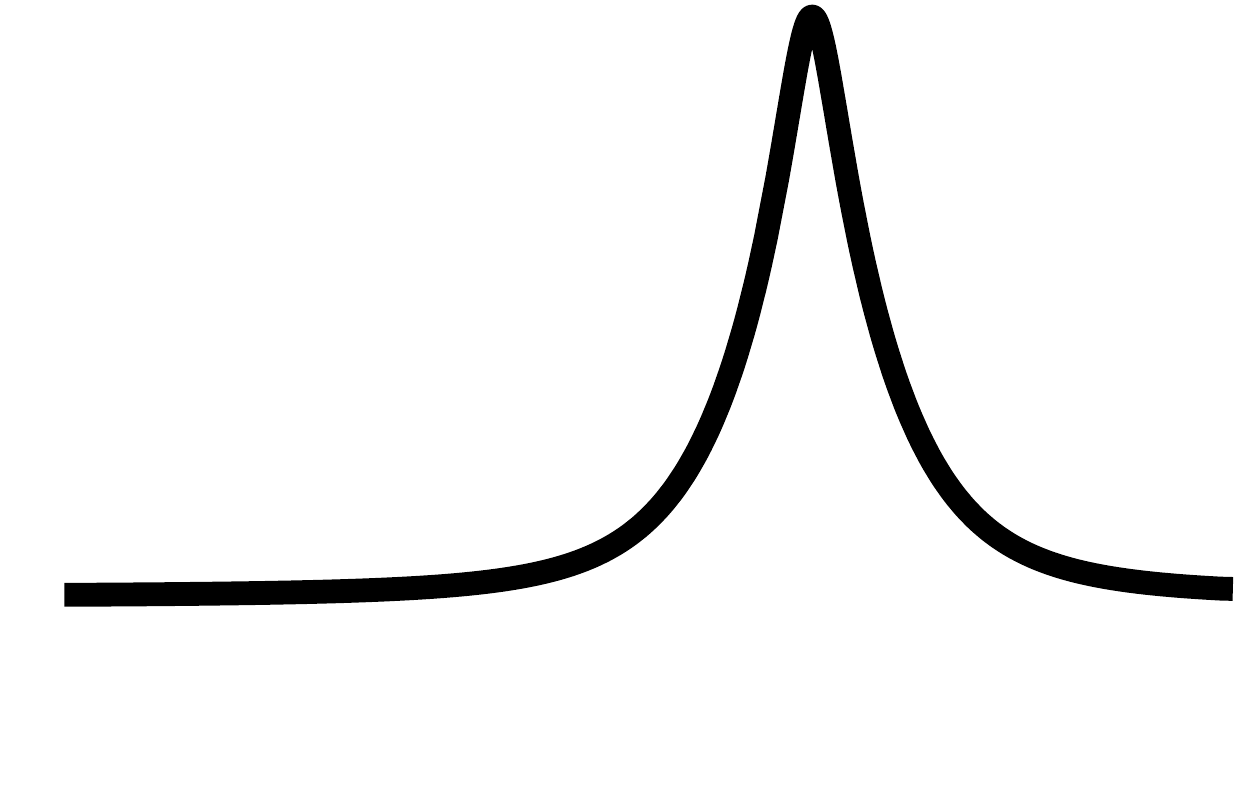}}\Big)_{f2}\,,\\[2ex]
\psi_{f2}^{(2)}(x) &= \Big(\raisebox{-9pt}{\includegraphics[width=.07\textwidth]{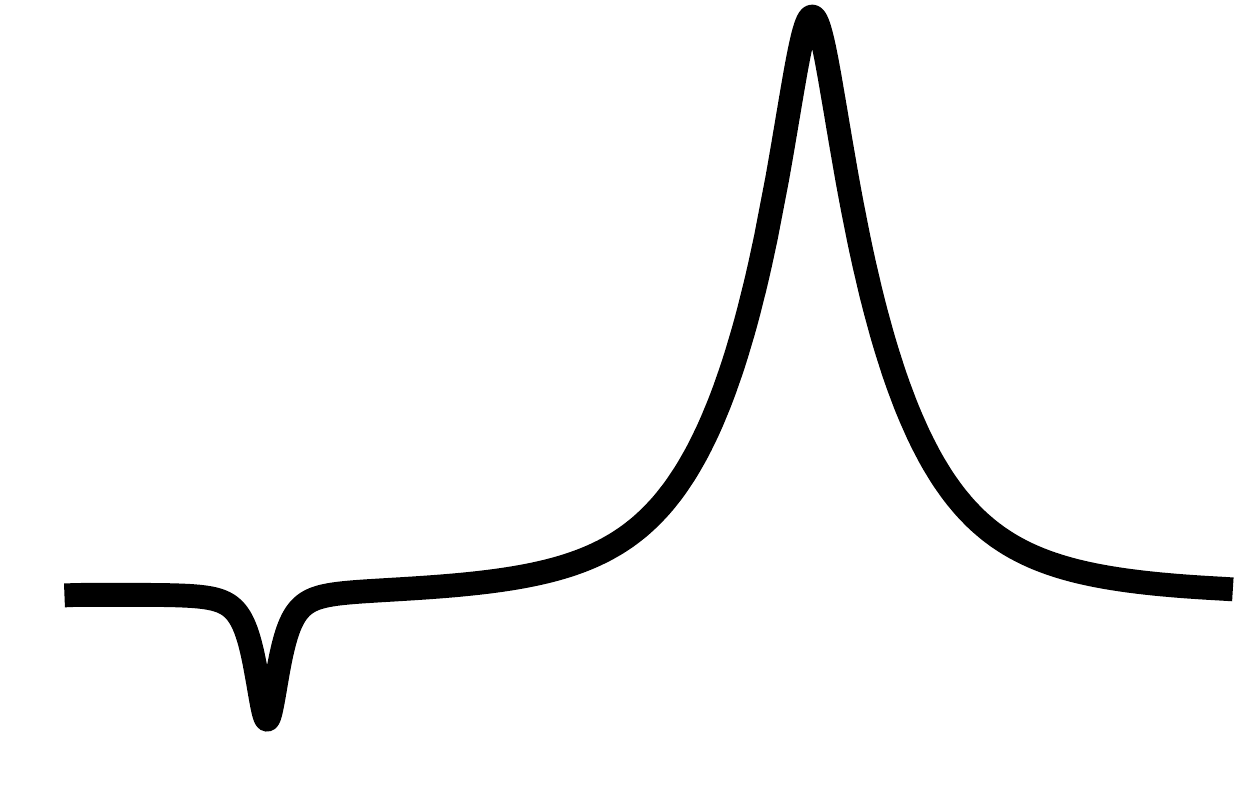}}\Big)_{f2}
= \Big(\raisebox{-9pt}{\includegraphics[width=.07\textwidth]{qzerosketch2a.pdf}}\Big)_{f2}- \mathbb{X}_2 \Big(\raisebox{-9pt}{\includegraphics[width=.07\textwidth]{qzerosketch1a.pdf}}\Big)_{f1}
\,,
\end{split}
\end{align}
where the left peak is located at $z_1$ and the right peak at $z_2$.

When the constituent-instantons are small, \Eq{eq:limit2},
the zero mode in \Eq{eq:psifi} becomes
\begin{align}\label{eq:psifi1}
\begin{split}
  \psi_{fi}(x,z_i) \approx \nu\, \frac{U_i \rho_i \gamma_\mu (x-z_i)_\mu}{|x-z_i|^4}\, \varphi_R
  = 2 \pi^2 \nu\, U_i \rho_i\, \Delta(x-z_i)\, \varphi_R \,,
\end{split}
\end{align}
where $\Delta(x-z)$ is the free propagator of a massless quark,
\begin{align}\label{eq:qprop}
\Delta(x-z) = \frac{\gamma_\mu (x-z)_\mu}{2 \pi^2 |x-z|^4}\,.
\end{align}
Hence, for $Q=2$ the quark zero modes reduce to a sum of free quark propagators and the overlap term,
\begin{align}
\psi_{f1}^{(2)}(x) \propto  U_1 \big[\rho_1\, \Delta(x-z_1) - \rho_2\, \mathbb{X}_1(x,z_1)\Delta(x-z_2)\big] \varphi_R\,.
\end{align}
This expression is essential in extracting the effective Lagrangian of quarks from the product of zero modes below.

Our results can be generalized to higher topological charge. It is less obvious how to
move away from the limit of small constituent-instantons, since then the terms generated in an effective Lagrangian involve derivatives of the quark fields, and so are of higher order in a derivative expansion.

\section{Generating functional for a small $Q$-instanton to leading order }\label{app:LOdilute}

We begin with the generating functional in a $Q$-instanton background, focusing on the determinant
of quark zero modes to leading order in the limit of small constituent-instantons. 
We evaluate the generating functional in the saddle point approximation to leading order,
where the stationary point is given by an instanton of topological charge $Q$.
This is natural, as (anti-) self-dual, topological gauge field configurations
are minima of the classical Yang-Mills action, assuming only that the action is finite \cite{Belavin:1975fg}.
The generating functional is
\begin{align}\label{eq:ZQapp}
Z^{\qq}[J] = \int\!\mathcal{D}\chi \exp\Big\{\!-S\big[\chi + \chi^{\qq}\big] + \int\!\! d^4x\, \bar\psi(x) J(x) \psi(x) \Big\} \, ,
\end{align}
where $\chi = (A_\mu, c, \bar c, \psi, \bar\psi)$ is the fluctuating field of gluons, ghosts and quarks,
$S[\chi]$ is the gauge-fixed action in Euclidean spacetime,
$\chi^{\qq} = (A_\mu^{\qq},0,0,0,0)$ is the $Q$-instanton background field, and $J$ a source for quarks.
To leading order in the saddle point approximation,
the action $S\big[\chi + \chi^{\qq}\big]$ is expanded in the $\chi^{\qq}$. The
linear terms vanish by the equations of motion, and the $\chi$ are integrated to quadratic order.
In the presence of the $Q$-instanton, all fields have zero modes related to the
invariance of the action under translations, dilatations and global gauge rotations.
There are $4 N_c |Q|$ collective coordinates describing their position ($z_i$), size ($\rho_i$),
and orientation in the gauge group ($U_i$). The set of these coordinates is denoted by $C_Q$.
Fluctuations in the directions of zero modes are large, so while the nonzero modes can be computed in the Gaussian approximation, the zero modes have to be treated exactly. To this end, one changes the integration over zero modes to an integration over collective coordinates, giving rise to a Jacobian $\mathcal{J}$.
This yields the $Q$-instanton density
\begin{align}
  n_Q(C_Q) \equiv {\rm e}^{- 8 \pi^2 |Q|/g^2} \; (\det{}_{\!\slashed{0}} \mathcal{M}_A)^{-1/2}
  (\det{}_{\!\slashed{0}} \mathcal{M}_c) (\det{}_{\!\slashed{0}} \slashed{D})\, (\det \mathcal{J}) \, ;
\end{align}
$\mathcal{M}_A$, $\mathcal{M}_c$, and $\mathcal{M}_\psi = \slashed{D}^{\qq}+J$ are the inverse
propagators of the gluons, ghosts, and quarks, respectively.
$\det{}_{\!\slashed{0}}$ denotes the determinant over nonzero modes.
To renormalize the contributions of large eigenvalues, it is understood that all nonzero-mode determinants are normalized with the determinant at vanishing gluon background field.
In this semi-classical approximation, the generating functional becomes
\begin{align}
Z^{\qq}[J] = \int\!dC_Q\, n_Q(C_Q) \det{}_{\!0}(J)\,.
\end{align}
$\det{}_{\!0}(J)$ is the determinant of the source $J$ in the space of quark zero modes.
This was first computed by 't Hooft in Ref. \cite{tHooft:1976rip} for $Q = 1$ and $N_c = 2$.
The generalization to $N_c \geq 3$ is given in Ref. \cite{Bernard:1979qt};
for $|Q| >1$, solutions are only known in certain limits, see {\it e.g.}\ Refs. \cite{Osborn:1981yf, Dorey:2002ik}.

For topological charge $Q$ at leading order in the limit of small constituent-instantons,
the gauge field is that of \Eq{eq:qismall}. The off-diagonal term in the quaternionic matrix $R$ in \Eq{eq:CRexpl} can be neglected and one can set $\xi_0 = 1$ in \Eq{eq:xi0exp}.
The quark zero modes in \Eq{eq:zerofull} then reduce to the $Q=1$ ones in \Eq{eq:psifi} and the zero mode determinant becomes
\begin{align}
\det{}_{\!0}(J) = \det \int\! d^4 x_{fi}\, \psi_{fi}^{\qq \dagger}(x_{fi}) J_{ij}^{fg}(x_{fi}) \psi_{gj}^{\qq}(x_{fi})
\approx \det \int\! d^4 x_{fi}\, \psi_{fi}^{\dagger}(x_{fi},z_i) J_{ij}^{fg}(x_{fi}) \psi_{gj}(x_{fi},z_i)\,,
\end{align}
where we do not sum over the indices.
The quark source $J$ is a $(N_f Q \times N_f Q)$-matrix in the space of zero modes, but
it suffices to consider a diagonal $J$, $J_{ii}^{ff} \equiv J_{fi}$. 
We will match the zero mode determinant to an effective multi-quark interaction, so the different contributions to the determinant can be obtained by permutation of the quark fields (cf.\ \Eq{eq:permdet}).
Using the explicit form of the quark zero modes in \Eq{eq:psifi},
the diagonal elements yield
\begin{align}\label{eq:dilQdet}
\begin{split}
\det{}_{\!0}(J)\Big|_\text{diagonal}
&= \prod_{i = 1}^Q \rho_i^{3N_f} \prod_{f = 1}^{N_f} \int\!d^4 x_{fi}\, \mathcal{F}_i(x_{fi}; z_i, \rho_i, U_i)\,,
 \end{split}
\end{align}
with the zero mode correlation function
\begin{align}\label{eq:F}
\mathcal{F}_i(x; z, \rho, U) 
= \frac{2}{\pi^2}\, \frac{U^\dagger \varphi_R^\dagger \gamma_\mu (x-z)_\mu \bar{J}_{i}(x)
 \gamma_\nu (x-z)_\nu \varphi_R U}{(x-z)^2 [(x-z)^2+\rho^2]^3} \,.
\end{align}
The quark zero modes have mass dimension two, while that of the source $J$ is zero.
We then introduce a source with canonical mass dimension,
\begin{align}\label{eq:jbar}
\bar J_{fi} = \rho_i^{-1} J_{fi}.
\end{align}
When $|x_{fi}-z_i| \gg \rho_i$, we can relate the zero modes to free quark propagators, \Eq{eq:psifi1}, and the quark zero mode correlations are related to those of free quarks,
\begin{align}\label{eq:Fprop}
\mathcal{F}_i(x;z, \rho, U) &\approx 8 \pi^2\, U^\dagger \varphi_R^\dagger \Delta(x-z) \bar{J}_{i}(x)
 \Delta(x-z) \varphi_R U\,.
\end{align}
Note that this is independent of the instanton scale $\rho$.
Since the generating functional involves and integral over
the collective coordinates, we can express \Eq{eq:dilQdet} as
\begin{align}\label{eq:zmno1}
\begin{split}
\det{}_{\!0}(J)\Big|_\text{diagonal}
&= \bigg[\rho_1^{3N_f} \int\!d^4 x_{f1}\, \mathcal{F}_i(x_{f1}; z_1, \rho_1, U_1) \bigg]^Q = \Big[\det{}_{\!0}(J)\Big|_{Q = 1}\Big]^Q\, .
 \end{split}
\end{align}
Thus to leading order for small constituent-instantons, the determinant over quark zero modes 
factorizes into a product of contributions from terms with $Q=1$.

As discussed in Refs. \cite{Bernard:1978ea, Osborn:1981yf, Dorey:2002ik},
the functional determinants of the gluons and ghosts factorize to order $\sim \rho^4/|R|^4$ 
for small constituent-instantons. Hence, the generating functional factorizes,
\begin{align}\label{eq:ZQdiluteapp}
  Z^{\qq}[J]\; \xrightarrow{\text{small}}\; \frac{1}{Q!}
  \bigg[\int\! dC_1\, n_1(\rho_1) \det{}_{\!0}(J)\Big|_{Q = 1}\bigg]^Q = \frac{1}{Q!}\big(Z^{(1)}[J]\big)^Q \, ,
\end{align}
where $Z^{(1)}[J]$ is the result for $Q = 1$.  
If ${\cal V}$ is the volume of space-time, this is $\sim {\cal V}^{|Q|}$, this is evidently
the expansion of the exponential of the term with $|Q| = 1$.  
Instead, what we need is a sub-leading term, proportional to a single power of ${\cal V}$.
We compute this term explicitly for $Q = 2$ in the next sections.

\section{Generating functional for $Q=2$ and one flavor}\label{app:Q2Nf1}

Here we compute the generating functional beyond leading order in the limit of small constituent-instantons for $Q=2$.
The determinant over the quark zero modes follow from \Eq{eq:qzero}.
We begin with the case of a single flavor and so drop the flavor index to obtain
\begin{align}\label{eq:1flavdet}
\begin{split}
\det{}_{\! 0}(J) &= 
\phantom{-}\int_{x_1, x_2} \bigg[\Big(\raisebox{-9pt}{\includegraphics[width=.07\textwidth]{qzerosketch1.pdf}}\Big)_{1}
 J_{1}(x_1)
  \Big(\raisebox{-9pt}{\includegraphics[width=.07\textwidth]{qzerosketch1.pdf}}\Big)_{1}\bigg]
\bigg[\Big(\raisebox{-9pt}{\includegraphics[width=.07\textwidth]{qzerosketch2.pdf}}\Big)_{2}
 J_{2}(x_2)
  \Big(\raisebox{-9pt}{\includegraphics[width=.07\textwidth]{qzerosketch2.pdf}}\Big)_{2}\bigg]\,.
\end{split}
\end{align}
We adopt a shorthand notation where $\int_{x_1} = \int d^4 x_1$, {\it etc.}
For small constituent-instantons, Eqs. \eq{eq:qzero} and \eq{eq:qzeropic} show that this contains
various contributions.   The integrations over the locations of the sources at $x_1$ and $x_2$ is always present,
and so we suppress it.  Instead we
concentrate on the integrals over the locations of the instantons, at $z_1$ and $z_2$.
Taking the dominant piece from each zero mode,
\begin{align}
\begin{split}
&\int_{z_1,z_2}\bigg[\Big(\raisebox{-9pt}{\includegraphics[width=.07\textwidth]{qzerosketch1a.pdf}}\Big)_{1} 
J_{1}(x_1) 
\Big(\raisebox{-9pt}{\includegraphics[width=.07\textwidth]{qzerosketch1a.pdf}}\Big)_{1}\bigg]
\bigg[\Big(\raisebox{-9pt}{\includegraphics[width=.07\textwidth]{qzerosketch2a.pdf}}\Big)_{2}
 J_{2}(x_2)
 \Big(\raisebox{-9pt}{\includegraphics[width=.07\textwidth]{qzerosketch2a.pdf}}\Big)_{2}\bigg]\\[1ex]
=&\int_{z_1} \Bigg\{  \bigg[\Big(\raisebox{-9pt}{\includegraphics[width=.07\textwidth]{qzerosketch1a.pdf}}\Big)_{1} 
J_{1}(x_1) 
\Big(\raisebox{-9pt}{\includegraphics[width=.07\textwidth]{qzerosketch1a.pdf}}\Big)_{1}\bigg] \Bigg\}
\int_{z_2}\Bigg\{  \bigg[\Big(\raisebox{-9pt}{\includegraphics[width=.07\textwidth]{qzerosketch2a.pdf}}\Big)_{2}
 J_{2}(x_2)
 \Big(\raisebox{-9pt}{\includegraphics[width=.07\textwidth]{qzerosketch2a.pdf}}\Big)_{2}\bigg]\Bigg\}\,.
\end{split}
\end{align}
This term has no overlap between the contributions at $z_1$ and $z_2$,
and so completely decomposes into two contributions from independent instantons with $|Q| = 1$.
These are $\sim {\cal V}^2$, as discussed in the previous section.

We need to extract local terms $\sim {\cal V}$, which are given by
\begin{align}\label{eq:ol1}
\begin{split}
&\int_{z_1,z_2}\bigg[\Big(\raisebox{-9pt}{\includegraphics[width=.07\textwidth]{qzerosketch1a.pdf}}\Big)_{1} 
J_{1}(x_1) 
\Big(\raisebox{-9pt}{\includegraphics[width=.07\textwidth]{qzerosketch1a.pdf}}\Big)_{1}\bigg]
\bigg[\mathbb{X}_2(x_2,z_2)\Big(\raisebox{-9pt}{\includegraphics[width=.07\textwidth]{qzerosketch1a.pdf}}\Big)_{2}
 J_{2}(x_2)
 \mathbb{X}_2(x_2,z_2)\Big(\raisebox{-9pt}{\includegraphics[width=.07\textwidth]{qzerosketch1a.pdf}}\Big)_{2}\bigg]\\[1ex]
 = &\int_{z_1} \Bigg\{ \bigg[\Big(\raisebox{-9pt}{\includegraphics[width=.07\textwidth]{qzerosketch1a.pdf}}\Big)_{1} 
J_{1}(x_1) 
\Big(\raisebox{-9pt}{\includegraphics[width=.07\textwidth]{qzerosketch1a.pdf}}\Big)_{1}\bigg]
\bigg[\Big(\raisebox{-9pt}{\includegraphics[width=.07\textwidth]{qzerosketch1a.pdf}}\Big)_{2}
 J_{2}(x_2)
 \Big(\raisebox{-9pt}{\includegraphics[width=.07\textwidth]{qzerosketch1a.pdf}}\Big)_{2}\bigg]\Bigg\}
 \int_{z_2}\Bigg\{  \mathbb{X}_2(x_2,z_2)^2\Bigg\}
 \,.
\end{split}
\end{align}
and
\begin{align}\label{eq:ol2}
\begin{split}
&\int_{z_1,z_2}\bigg[ \mathbb{X}_1(x_1,z_1) \Big(\raisebox{-9pt}{\includegraphics[width=.07\textwidth]{qzerosketch2a.pdf}}\Big)_{1} 
J_{1}(x_1) 
\mathbb{X}_1(x_1,z_1) \Big(\raisebox{-9pt}{\includegraphics[width=.07\textwidth]{qzerosketch2a.pdf}}\Big)_{1}\bigg]
\bigg[\Big(\raisebox{-9pt}{\includegraphics[width=.07\textwidth]{qzerosketch2a.pdf}}\Big)_{2}
 J_{2}(x_2)
 \Big(\raisebox{-9pt}{\includegraphics[width=.07\textwidth]{qzerosketch2a.pdf}}\Big)_{2}\bigg]\\[1ex]
 = &\int_{z_1} \Bigg\{  \mathbb{X}_1(x_1,z_1)^2\Bigg\}
\int_{z_2} \Bigg\{ \bigg[\Big(\raisebox{-9pt}{\includegraphics[width=.07\textwidth]{qzerosketch2a.pdf}}\Big)_{1} 
J_{1}(x_1) 
\Big(\raisebox{-9pt}{\includegraphics[width=.07\textwidth]{qzerosketch2a.pdf}}\Big)_{1}\bigg]
\bigg[\Big(\raisebox{-9pt}{\includegraphics[width=.07\textwidth]{qzerosketch2a.pdf}}\Big)_{2}
 J_{2}(x_2)
 \Big(\raisebox{-9pt}{\includegraphics[width=.07\textwidth]{qzerosketch2a.pdf}}\Big)_{2}\bigg]\Bigg\}\,.
\end{split}
\end{align}
These terms are given by four $Q = 1$ zero modes centered around a single $z_i$. The integral
\begin{align}
\mathcal{I}_{1,i} = \int_{z_i} \mathbb{X}_i(x_i,z_i)^2\, ,
\end{align}
represents the ``leakage'' from the constituent-instanton at $z_2$ to $z_1$, or {\it vice versa}.
This integral can be done analytically,
\begin{align}\label{eq:OI1}
\mathcal{I}_{1,i}(\rho_1,\rho_2) 
 = \rho_1^2 \rho_2^2  \int\!d^4 z_i\, \frac{(x_i-z_i)^2}{[(x_i-z_i)^2+\rho_i^2]^{3}} 
  =\pi^2 \rho_1^2 \rho_2^2 \bigg[
  \ln\!\bigg(\frac{R_0^2}{R_0^2+ \rho_i^2}\bigg)
  - \frac{R_0^2(3 R_0^2 + 2 \rho_i^2)}{2 (R_0^2 + \rho_i^2)^2}
   \bigg] \; .
\end{align}
For a single flavor this overlap integral is dominated by large distances, $|x-z_i|$.
There is a logarithmic divergence in the infrared, which we cutoff at a scale $R_0 \gg \rho_i$.
Presumably $R_0$ is related to the average separation between instantons and anti-instantons in the vacuum
\cite{Shuryak:1978yk,Diakonov:1983hh,Diakonov:2009jq}.

With this, \Eq{eq:ol1} becomes
\begin{align}\label{eq:zmo1}
\begin{split}
& \mathcal{I}_{1,2} \int_{z_1} \bigg[\Big(\raisebox{-9pt}{\includegraphics[width=.07\textwidth]{qzerosketch1a.pdf}}\Big)_{1} 
J_{1}(x_1) 
\Big(\raisebox{-9pt}{\includegraphics[width=.07\textwidth]{qzerosketch1a.pdf}}\Big)_{1}\bigg]
\bigg[\Big(\raisebox{-9pt}{\includegraphics[width=.07\textwidth]{qzerosketch1a.pdf}}\Big)_{2}
 J_{2}(x_2)
 \Big(\raisebox{-9pt}{\includegraphics[width=.07\textwidth]{qzerosketch1a.pdf}}\Big)_{2}\bigg]\\[1ex]
 &\quad = \mathcal{I}_{1,2} \int_{z_1} \Big[\psi_{11}^\dagger(x_1,z_1) J_{1}(x_1) \psi_{11}(x_1,z_1)\Big]
 \Big[\widehat{\psi}_{11}^{\,\dagger}(x_2,z_1) J_{2}(x_2) \widehat{\psi}_{11}(x_2,z_1)\Big]\\[1ex]
&\quad= \mathcal{I}_{1,2} \int_{z_1}
\Big[ \rho_1^3\, \mathcal{F}_1(x_1; z_1, \rho_1, U_1) \Big]
\Big[\rho_1^2 \rho_2\, \mathcal{F}_2(x_2; z_1, \rho_1, U_2) \Big]\,,
\end{split}
\end{align}
where we used Eqs.\ \eq{eq:qzero}, \eq{eq:F} and \eq{eq:jbar}. For \Eq{eq:ol2} we find the same result, but with the charge indices ${}_1$ and ${}_2$ interchanged.
The peculiar powers of the instanton sizes in \Eq{eq:zmo1}, as compared to the leading-order result in \Eq{eq:zmno1}, arise from the overlap of the zero modes. The contributions of the first zero mode, $\psi_{11}^{(2)}$, and the second zero mode, $\psi_{12}^{(2)}$, centered around the same point $z_1$, each have the scale $\rho_1$ so that they each contribute $\rho_1^2$ to the determinant. Rescaling the two quark sources $J_1$ and $J_2$ according to \Eq{eq:jbar} yields additional factors $\rho_1$ and $\rho_2$ respectively. For a single quark flavor, the local quark zero mode determinant therefore goes like $\rho_1^5 \rho_2$, instead of $\rho_1^3 \rho_2^3$ for the non-local contribution at leading order.

We have shown that the overlap between the different quark zero modes arises only beyond leading order in the limit of small constituent instantons. This is essential for deriving a local effective action $\sim \mathcal{V}$. Remarkably, even to order $\zeta^3$, the gauge contribution to the path integral factorizes and we can still use \Eq{eq:qismall} for the 2-instanton \cite{Bernard:1978ea}.
So only corrections of $\mathcal{O}(\zeta^3)$ for the quark zero mode determinant need to be included.
With the results above, the local part of the $Q=2$ partition function $Z^{(2)}[J]$, \Eq{eq:ZQdilute}, for $N_f = 1$ reduces to
\begin{align}\label{eq:nf1Z2a}
Z^{(2)}[J]\Big|_\text{local} =
 \frac{1}{2} \int\! d\rho_1\, n_1(\rho_1) \rho_1^5
 \int\! d\rho_2\, n_1(\rho_2) \rho_2\, \mathcal{I}_{1,2}
\int\!d^4z_1 \prod_{i=1}^{2}\bigg\{\int\!dU_i\int\!d^4x_i\, \mathcal{F}_i(x_i;z_1,U_i)\bigg\}
+ (1\leftrightarrow 2)\,.
\end{align}
Since this expression is symmetric under the exchange of
the topological charge indices ${}_1$ and ${}_2$, we finally arrive at
\begin{align}\label{eq:nf1Z2}
Z^{(2)}[J]\Big|_\text{local} &=
 \int\! d\rho_1\, n_1(\rho_1) \rho_1^5
 \int\! d\rho_2\, n_1(\rho_2) \rho_2\, \mathcal{I}_{1,2}
\int\!d^4z_1 \prod_{i=1}^{2}\bigg\{\int\!dU_i\int\!d^4x_i\, \mathcal{F}_i(x_i;z_1,U_i)\bigg\}\,,
\end{align}
where the $Q=1$ instanton density $n_1$ is given by \Eq{instanton_density}.

\section{Generating functional for $Q=2$ and $N_f \geq 2$}\label{app:Q2Nf}

The discussion for two or more flavors
is a straightforward generalization of that for a single flavor.
Taking the quark source to be diagonal, the determinant is
\begin{align}\label{eq:2flavdet}
\begin{split}
\det{}_0(J)\Big|_{\text{diag}} &= \prod_{f = 1}^{N_f}\Bigg\{
\int_{x_{f1}} \bigg[\Big(\raisebox{-9pt}{\includegraphics[width=.07\textwidth]{qzerosketch1.pdf}}\Big)_{f1}
 J_{f1}(x_{f1})
  \Big(\raisebox{-9pt}{\includegraphics[width=.07\textwidth]{qzerosketch1.pdf}}\Big)_{f1}\bigg]
  \int_{x_{f2}} \bigg[\Big(\raisebox{-9pt}{\includegraphics[width=.07\textwidth]{qzerosketch2.pdf}}\Big)_{f2}
 J_{f2}(x_{f2})
  \Big(\raisebox{-9pt}{\includegraphics[width=.07\textwidth]{qzerosketch2.pdf}}\Big)_{f2}\bigg]
  \Bigg\}\, .
\end{split}
\end{align}
There are numerous contributions $\sim {\cal V}^2$.
Assuming that the instantons are uncorrelated, the leading term is
\begin{align}
\int_{z_1} \prod_{f = 1}^{N_f}\Bigg\{  \bigg[\Big(\raisebox{-9pt}{\includegraphics[width=.07\textwidth]{qzerosketch1a.pdf}}\Big)_{f1} 
J_{f1}(x_{f1}) 
\Big(\raisebox{-9pt}{\includegraphics[width=.07\textwidth]{qzerosketch1a.pdf}}\Big)_{f1}\bigg] \Bigg\}
\int_{z_2}\prod_{f = 1}^{N_f} \Bigg\{ \bigg[\Big(\raisebox{-9pt}{\includegraphics[width=.07\textwidth]{qzerosketch2a.pdf}}\Big)_{f2}
 J_{f2}(x_{f2})
 \Big(\raisebox{-9pt}{\includegraphics[width=.07\textwidth]{qzerosketch2a.pdf}}\Big)_{f2}\bigg]\Bigg\}\, ,
\end{align}
as discussed in \App{app:LOdilute}.

The terms of interest, $\sim {\cal V}$, are given by
\begin{align}\label{eq:twoflavloc1}
\int_{z_1}  \prod_{f = 1}^{N_f}\Bigg\{  \bigg[\Big(\raisebox{-9pt}{\includegraphics[width=.07\textwidth]{qzerosketch1a.pdf}}\Big)_{f1} 
J_{f1}(x_{f1}) 
\Big(\raisebox{-9pt}{\includegraphics[width=.07\textwidth]{qzerosketch1a.pdf}}\Big)_{f1}\bigg]
\bigg[\Big(\raisebox{-9pt}{\includegraphics[width=.07\textwidth]{qzerosketch1a.pdf}}\Big)_{f2}
 J_{f2}(x_{f2})
 \Big(\raisebox{-9pt}{\includegraphics[width=.07\textwidth]{qzerosketch1a.pdf}}\Big)_{f2}\bigg]\Bigg\}
\int_{z_2}  \prod_{f = 1}^{N_f}\Bigg\{  \mathbb{X}_2(x_{f2},z_2)^2\Bigg\}\,,
\end{align}
and
\begin{align}\label{eq:twoflavloc2}
\int_{z_1}  \prod_{f = 1}^{N_f}\Bigg\{  \mathbb{X}_1(x_{f1},z_1)^2\Bigg\}
\int_{z_2}  \prod_{f = 1}^{N_f}\Bigg\{  \bigg[\Big(\raisebox{-9pt}{\includegraphics[width=.07\textwidth]{qzerosketch2a.pdf}}\Big)_{f1} 
J_{f1}(x_{f1}) 
\Big(\raisebox{-9pt}{\includegraphics[width=.07\textwidth]{qzerosketch2a.pdf}}\Big)_{f1}\bigg]
\bigg[\Big(\raisebox{-9pt}{\includegraphics[width=.07\textwidth]{qzerosketch2a.pdf}}\Big)_{f2}
 J_{f2}(x_{f2})
 \Big(\raisebox{-9pt}{\includegraphics[width=.07\textwidth]{qzerosketch2a.pdf}}\Big)_{f2}\bigg]\Bigg\}\,.
\end{align}
Using Eqs.\ \eq{eq:qzero}, \eq{eq:F} and \eq{eq:jbar}, the integral over $z_1$ in \Eq{eq:twoflavloc1} equals
\begin{align}
\begin{split}
&\int_{z_1}  \prod_{f = 1}^{N_f}\Bigg\{  \bigg[\Big(\raisebox{-9pt}{\includegraphics[width=.07\textwidth]{qzerosketch1a.pdf}}\Big)_{f1} 
J_{f1}(x_{f1}) 
\Big(\raisebox{-9pt}{\includegraphics[width=.07\textwidth]{qzerosketch1a.pdf}}\Big)_{f1}\bigg]
\bigg[\Big(\raisebox{-9pt}{\includegraphics[width=.07\textwidth]{qzerosketch1a.pdf}}\Big)_{f2}
 J_{f2}(x_{f2})
 \Big(\raisebox{-9pt}{\includegraphics[width=.07\textwidth]{qzerosketch1a.pdf}}\Big)_{f2}\bigg]\Bigg\}\\[1ex]
 &\quad = \int_{z_1} \prod_{f=1}^{N_f} \Big[\psi_{f1}^\dagger(x_{f1},z_1) J_{1}(x_{f1}) \psi_{f1}(x_{f1},z_1)\Big]
 \Big[\widehat{\psi}_{f1}^{\,\dagger}(x_{f2},z_1) J_{2}(x_{f2}) \widehat{\psi}_{f1}(x_{f2},z_1)\Big]\\[1ex]
&\quad= \int_{z_1} \prod_{f=1}^{N_f} \bigg\{
\Big[ \rho_1^3\, \mathcal{F}_1(x_{f1}; z_1, \rho_1, U_1) \Big]
\Big[\rho_1^2 \rho_2\, \mathcal{F}_2(x_{f2}; z_1, \rho_1, U_2) \Big]\bigg\}\,,
\end{split}
\end{align}
and similarly for \Eq{eq:twoflavloc2}.
The overlap integral for any $N_f$ is given by:
\begin{align}
  \mathcal{I}_{N_f,i}(\rho_1, \rho_2, \{x_{fi}\}) = \int_{z_i}  \prod_{f = 1}^{N_f}
  \mathbb{X}_i(x_{fi},z_i)^2 = (\rho_1 \rho_2)^{2 N_f}  \int\!d^4 z_i
  \prod_{f = 1}^{N_f} \frac{(x_{fi}-z_i)^2}{[(x_{fi}-z_i)^2+\rho_i^2]^{3}}\,.
\end{align}
In general this integral depends on the locations of the two sources.  
For an instanton with $Q=2$, though, if the constituent-instantons are small, then
the zero modes $\psi_{f2}^{(2)}(x)$ generate
the overlap term in \Eq{eq:twoflavloc1}. This overlap stems from configurations where
$\psi_{f2}^{(2)}(x_{f2})$ is closer the first constituent-instanton, {\it i.e.}\
$|x_{f2}-z_1| \ll |x_{f2}-z_2| \approx |z_1- z_2| = |R_{12}|$.  Then
\begin{align}
  \psi_{f2}^{(2)}(x_{f2})\Big|_{|x_{f2}-z_1| \ll |x_{f2}-z_2|} \approx - \widehat{\psi}_{f1}(x_{f2},z_1)
  \frac{\rho_1\rho_2\, |R_{12}|}{(R_{12}^2+\rho_2^2)^{3/2}}\,.
\end{align}
This limit is consistent with Eqs.\ \eq{eq:limit1} and \eq{eq:limit2} as long as $|x_{f2}-z_1| \gg \rho_1, \rho_2$.
Other terms $\sim {\cal V}^2$ are dominated by configurations
where at least one of the zero modes $\psi_{f2}^{(2)}(x_{f2})$ is close to $z_2$,
and are suppressed for $|x_{f2}-z_1| \ll |x_{f2}-z_2|$. 
The analogous statement is true for the overlap from $\psi_{f1}^{(2)}(x_{f1})$ in \Eq{eq:twoflavloc2}.
Hence, in this case the overlap term only depends on the instanton size and the distance between the instantons,
\begin{align}
\mathbb{X}_i = \frac{\rho_1\rho_2\, |R_{12}|}{(R_{12}^2 + \rho_i^2)^{3/2}}\,,
\end{align}
and the quark zero mode determinant is dominated by a term $\sim {\cal V}$.
The overlap integral for $N_f \geq 2$ becomes
\begin{align}\label{eq:OINf}
\begin{split}
\mathcal{I}_{N_f,i}(\rho_1,\rho_2) &= (\rho_1 \rho_2)^{2 N_f}  \int\!d^4 R_{12} \prod_{f = 1}^{N_f} \frac{R_{12}^2}{(R_{12}^2+\rho_i^2)^{3}}
= (\rho_1 \rho_2)^{2 N_f}  \int\!d^4 R_{12}\, \frac{R_{12}^{2 N_f}}{(R_{12}^2+\rho_i^2)^{3 N_f}}\\
&=  \frac{\pi^2 (N_f+1)! (2 N_f - 3)!}{(3N_f-1)!} \, \left(\frac{\rho_1 \rho_2}{\rho_i^2}\right)^{2 N_f} \rho_i^4\,.
\end{split}
\end{align}
For any $N_f$, the partition function for $Q = 2$ is
\begin{align}\label{eq:nfZ2}
\begin{split}
Z^{(2)}[J]\Big|_\text{local} &=
\int\! d\rho_1\, n_1(\rho_1) \rho_1^{5 N_f}
\int\! d\rho_2\, n_1(\rho_2) \rho_2^{N_f}\, \mathcal{I}_{N_f,2}
\int\!d^4z_1 \prod_{i=1}^{2}\bigg\{\int\!dU_i \prod_{f = 1}^{N_f}\int\!d^4x_{fi}\, \mathcal{F}_i(x_{fi};z_1,U_i)\bigg\}\,,
\end{split}
\end{align}
with $\mathcal{F}_i$ defined in \Eq{eq:F}.
We emphasize that since at large distances $\mathcal{F}_i$ contains two free quark propagators \eq{eq:Fprop},
the generating functional is that of a correlation function between $2 N_f Q$ quarks.

\section{The effective interaction}

We now derive the effective action from the quark zero mode determinant computed in the previous sections.
The main trick is to exploit the fact that far away from the instanton,
the determinant can be expressed in terms of free quark propagators, cf.\ \Eq{eq:Fprop}, so that 
it mimics an operator without a background field, located at the position of the instanton.

\subsection{Any topological charge at leading order}

Before we discuss the local interaction for $Q = 2$, we consider the result to leading order for a small
$Q$-instanton, under the ansatz:
\begin{align}\label{eq:ZQeffansatz}
\begin{split}
Z_\text{eff,LO}^{\qq+}[\bar J] &= \int\! \mathcal{D}\psi \mathcal{D}\bar\psi \exp\bigg\{\!-S[\chi] + \int\!\! d^4x\, \bar\psi \bar J \psi \bigg\} \,V_\text{eff,LO}^{\qq+}\,,\\[2ex]
V_\text{eff,LO}^{\qq+} &=  \frac{\kappa_{Q, \text{LO}}}{K_{Q,N_f}} \prod_{i=1}^{Q} \Bigg\{ \int\! d^4z_i \int\! dU_i \prod_{f=1}^{N_f} \big[\bar\psi_f(z_i)\, \omega_i \big] \big[\bar \omega_i\, \psi_f(z_i) \big] \Bigg\}\, ;
\end{split}
\end{align}
where again without loss of generality we assume $Q$ is positive. This ansatz applies only to leading order (LO) for small
constituent-instantons.  $\omega_i$ are constant tensors carrying spin and color which are determined below,
and $K_{Q,N_f}$ is defined in \Eq{eq:KQNf}.
The pre-exponential factor $V_\text{eff,LO}^{\qq+}$ generates a non-local
$2 N_f Q$-correlation function with coupling strength $\kappa_Q$.
The superscript ${}^+$ indicates that this is contribution from instantons; ${}^-$ denotes that from anti-instantons.
Because of Fermi statistics, this term can be rewritten as a determinant,
\begin{align}\label{eq:permdet}
\prod_{f=1}^{N_f} (\bar \psi_f \omega_i)(\bar\omega_i \psi_f) = \frac{1}{N_f!} \det_{fg}\big[(\bar \psi_f \omega_i)(\bar\omega_i \psi_g) \big]\,.
\end{align}
This explains why we could take the quark source $J$ to be diagonal in color and flavor, as
all other contributions are given by permutations of the quark fields.
The correlation function generated by $V_\text{eff,LO}^{\qq+}$
can be computed by expressing the exponential as a power series in $\bar J$,
\begin{align}\label{eq:sourceexp}
\nonumber
  e^{-S + \int_x \bar \psi \bar J \psi} = e^{-S} \Bigg\{1 &+ \int_{x_{11}}\, \bar\psi(x_{11}) \bar J(x_{11}) \psi(x_{11})
+ \frac{1}{2} \int_{x_{11}} \, \bar\psi(x_{11}) \bar J(x_{11}) \psi(x_1)\; \int_{x_{12}}\,
\bar\psi(x_{12}) \bar J(x_{12}) \psi(x_{12})\\[1ex]
&+ \dots + \frac{1}{(N_f Q)!} \prod_{i=1}^{Q}
\prod_{f=1}^{N_f} \int_{x_{fi}}\, \bar\psi(x_{fi}) \bar J(x_{fi}) \psi(x_{fi}) + \dots \Bigg\}\, .
\end{align}
Wick's theorem is then used to contract the quarks from the sources with those
in $V_\text{eff,LO}^{\qq+}$. Note our suggestive notation for the vertex locations
in \Eq{eq:ZQeffansatz} and the locations of the sources in \Eq{eq:sourceexp}.
For small constituent-instantons, the $z_i$ in \Eq{eq:ZQeffansatz}
are by definition widely separated. As a result, contractions of quark fields are suppressed
except for those where all quarks sourced at $x_{fi}$ are contracted at $z_j$ in $V_\text{eff,LO}^{\qq+}$.
Any other contraction involves at least one propagator $\Delta(z_i,z_j)$, with $i \neq j$,
which is suppressed for small constituent-instantons.
Hence, only the term of order $N_f Q$ in $\bar J$ in \Eq{eq:sourceexp} contributes.
For fixed $f$, there are $Q!$ equivalent ways to contract the quarks at $x_{fi}$ with
the ones at $z_j$. Since this can be done for each $f$,
there are $(Q!)^{N_f}$ equivalent contributions. All other contractions are suppressed,
since they contain at least one $\Delta(z_i,z_j)$.
We combine this into the combinatorial factor
\begin{align}\label{eq:KQNf}
K_{Q,N_f} = \frac{(Q!)^{N_f}}{(N_f Q)!}\,.
\end{align}
Expanding the exponential in powers of the the
sources and using Wick's theorem for small constituent-instantons, $Z_\text{eff,LO}^{\qq+}[\bar J]$
is dominated by a $2 N_f Q$-quark correlation function multiplied by $K_{Q,N_f}$.
To compensate for this factor, we introduced a
factor of $1/K_{Q,N_f}$ for the effective coupling in \Eq{eq:ZQeffansatz}.
In all we find
\begin{align}\label{eq:ZeffQcorr}
\begin{split}
  &\Bigg\langle\frac{1}{(N_f Q)!} \prod_{i=1}^{Q} \prod_{f=1}^{N_f}\bigg[\int_{x_{fi}}\,
  \bar\psi(x_{fi}) \bar J(x_{fi}) \psi(x_{fi})\bigg]\,
 V_\text{eff,LO}^{\qq+}\Bigg\rangle  \\[1ex]
&\quad = \kappa_{Q,\text{LO}}\, \prod_{i=1}^Q \int_{z_i} \int\!dU_i\, \prod_{f=1}^{N_f} \int_{x_{fi}}\,
\bar\omega_i\, \Delta(x_{fi}-z_i)\, \bar J (x_{fi})\, \Delta(x_{fi}-z_i)\, \omega_i\,.
\end{split}
\end{align}
Comparing this with \Eq{eq:ZQdiluteapp}, using \Eq{eq:Fprop}, we find
\begin{align}\label{eq:QLOcompare}
\begin{split}
&\kappa_{Q,\text{LO}}\, \prod_{i=1}^Q \int\!d^4z_i \int\!dU_i\, \prod_{f=1}^{N_f} \int\!d^4x_{fi}\,
\bar\omega_i\, \Delta(x_{fi}-z_i)\, \bar J (x_{fi})\, \Delta(x_{fi}-z_i)\, \omega_i \\[1ex]
&\quad=\frac{(8\pi^2)^{N_f Q}}{Q!} \bigg[\int\!d\rho_i\, n_1(\rho_i) \rho_i^{3N_f}\bigg]^Q
\prod_{i=1}^Q\int\!d^4z_i \int\!dU_i\, \prod_{f=1}^{N_f} \int\!d^4x_{fi}\,
U_i^\dagger \varphi_R^\dagger \Delta(x_{fi}-z_i) \bar{J}(x_{fi}) \Delta(x_{fi}-z_i) \varphi_R U_i\,,
\end{split}
\end{align}
From this, the effective coupling is
\begin{align}\label{eq:kappaQLO}
\kappa_{Q,\text{LO}} = \frac{1}{Q!} \bigg[(8\pi^2)^{N_f} \int\!d\rho_i\, n_1(\rho_i) \rho_i^{3N_f}\bigg]^Q\,.
\end{align}
Aside from the combinatoric factor, this is precisely the effective coupling
derived in \cite{tHooft:1976rip, tHooft:1976snw, tHooft:1986ooh} for the single instanton to the $Q^{\rm th}$
power. From the integrands on both sides of \Eq{eq:QLOcompare} we infer that the tensor $\omega$ obeys the identity,
\begin{align}
(\omega_i)_\alpha^a\, (\bar\omega_i)_\beta^a = \varphi_R^{\alpha b} U_i^{ba} (U_i^\dagger)^{ac} (\varphi_R^\dagger)^{c \beta}\,,
\end{align}
where the color indices ($a,\, b,\, c$) and spinor indices ($\alpha,\, \beta$) are explicit here.
The color structure of $\omega_i$ is fixed by requiring that it carries the global color orientation $U_i$,
\begin{align}
 (U_i)^{ab} \omega_\alpha^b = (\omega_i)_\alpha^a\,.
\end{align}
Further, from the explicit form of the spinor $\varphi_R$ in \Eq{eq:spinorR}
\begin{align}
\varphi_R^{\alpha a} (\varphi_R^\dagger)^{a \beta} = \mathbb{P}_R^{\alpha\beta}\,,
\end{align}
where $\mathbb{P}_R$ is the right-handed projection operator defined in \Eq{eq:chiproj}. This implies
\begin{align}\label{eq:omegaid}
\omega_\alpha^a\, \bar\omega_\beta^a = \mathbb{P}_R^{\alpha\beta}\,.
\end{align}
The integration over the orientation in the gauge group in \Eq{eq:ZQeffansatz} can now be carried out.
Since we have the same integral for different topological charge indices $i$,
the integral for fixed $i$ is done following Refs.
\cite{tHooft:1976rip, tHooft:1976snw, tHooft:1986ooh}.
The final result is this result to the $Q^{\rm th}$ power,
\begin{align}\label{eq:colint1}
\int\! dU_i \prod_{f=1}^{N_f} \big[\bar\psi_f(z_i)\, \omega_i \big] \big[\bar \omega_i\, \psi_f(z_i) \big] = \int\! dU_i \prod_{f=1}^{N_f} \big[\bar\psi_f(z_i)\, U_i\, \omega \big] \big[\bar \omega\, U_i^\dagger\, \psi_f(z_i) \big]\,.
\end{align}
For a $SU(N_c)$ gauge group
\begin{align}
\int\! dU_i \prod_{f=1}^{N_f} U_i^{a_f b_f} (U_i^\dagger)^{c_f d_f}\,,
\end{align}
where $U_i$ is an element of $SU(N_c)/\mathcal{I}_{N_c}$, with $\mathcal{I}_{N_c}$ the stability group of the instanton,
the set of $SU(N_c)$-transformations that leave the instanton unchanged.
For $N_c = 2$ this is the identity, and $dU_i$ is the corresponding Haar measure.
For arbitrary $N_f$ and $N_c$ he integration is more involved.
In general, the group integration in \Eq{eq:colint1} yields products of terms which are color singlet,
$\sim \overline{\psi}_L \psi_R$, and non-singlet, $\sim \psi_L \psi_L$.
In vacuum only the color singlet terms, $\sim \overline{\psi}_L \psi_R$, are important, but at nonzero
density, the non-singlet terms $\sim \psi_L \psi_L$ affect color superconductivity.
For two flavors, product of color singlet terms is extracted from
\begin{align}\label{eq:uintnf2}
\int\! dU_i\, U_i^{a_1 b_1} (U_i^\dagger)^{c_1 d_1} U_i^{a_2 b_2} (U_i^\dagger)^{c_2 d_2} = c_{N_c} \big(\delta^{a_1 d_1}\delta^{a_2 d_2}\delta^{b_1 c_1}\delta^{b_2 c_2}
+ \delta^{a_1 d_2}\delta^{a_2 d_1}\delta^{b_1 c_2}\delta^{b_2 c_1} \big) + (\text{non-singlet})\, ;
\end{align}
$c_{N_c}$ is a $N_c$-dependent constant. We find
\begin{align}
&\int\! dU_i \prod_{f=1}^{N_f} \big[\bar\psi_f(z_i)\, U_i\, \omega \big] \big[\bar \omega\, U_i^\dagger\, \psi_f(z_i) 
\big]\bigg|_\text{singlet}\\[1ex] \nonumber
&\quad= c_{N_c} \big(\delta^{a_1 d_1}\delta^{a_2 d_2}\delta^{b_1 c_1}\delta^{b_2 c_2}
+ \delta^{a_1 d_2}\delta^{a_2 d_1}\delta^{b_1 c_2}\delta^{b_2 c_1} \big) 
\Big[\bar\psi_1^{\alpha a_1}(z_i)\, \omega_\alpha^{b_1}\, \bar \omega_\beta^{c_1}\, \psi_1^{\beta d_1}(z_i)\,
\bar\psi_2^{\alpha a_2}(z_i)\, \omega_\alpha^{b_2}\, \bar \omega_\beta^{c_2}\, \psi_1^{\beta d_2}(z_i) \Big]\\[1ex] \nonumber
&\quad= c_{N_c}\Big[
\bar\psi_1^{\alpha a_1}(z_i)\, \omega_\alpha^{b_1}\, \bar \omega_\beta^{b_1}\, \psi_1^{\beta a_1}(z_i) \,
\bar\psi_2^{\alpha a_2}(z_i)\, \omega_\alpha^{b_2}\, \bar \omega_\beta^{b_2}\, \psi_2^{\beta a_2}(z_i)
+
\bar\psi_1^{\alpha a_1}(z_i)\, \omega_\alpha^{b_1}\, \bar \omega_\beta^{b_2}\, \psi_1^{\beta a_2}(z_i)\,
\bar\psi_2^{\alpha a_2}(z_i)\, \omega_\alpha^{b_2}\, \bar \omega_\beta^{b_1}\, \psi_2^{\beta a_1}(z_i)
\Big]\\[1ex] \nonumber
&\quad= c_{N_c}\Big[
\bar\psi_1^{\alpha a_1}(z_i)\, \omega_\alpha^{b_1}\, \bar \omega_\beta^{b_1}\, \psi_1^{\beta a_1}(z_i) \,
\bar\psi_2^{\alpha a_2}(z_i)\, \omega_\alpha^{b_2}\, \bar \omega_\beta^{b_2}\, \psi_2^{\beta a_2}(z_i)
-
\bar\psi_1^{\alpha a_1}(z_i)\, \omega_\alpha^{b_1}\, \bar \omega_\beta^{b_1}\, \psi_2^{\beta a_1}(z_i)\,
\bar\psi_2^{\alpha a_2}(z_i)\, \omega_\alpha^{b_2}\, \bar \omega_\beta^{b_2}\, \psi_1^{\beta a_2}(z_i)
\Big]\,.
\end{align}
Now we can apply the identity for $\omega$ in \Eq{eq:omegaid} to arrive at,
\begin{align}
\begin{split}
&\int\! dU_i \prod_{f=1}^{N_f} \big[\bar\psi_f(z_i)\, U_i\, \omega \big] \big[\bar \omega\, U_i^\dagger\, \psi_f(z_i) 
\big]\bigg|_\text{singlet}\\[1ex]
&\quad= c_{N_c}\Big[
\bar\psi_1(z_i)\, \mathbb{P}_R\, \psi_1(z_i) \,
\bar\psi_2(z_i)\, \mathbb{P}_R\, \psi_2(z_i)
-
\bar\psi_1(z_i)\, \mathbb{P}_R\, \psi_2(z_i)\,
\bar\psi_2(z_i)\,\mathbb{P}_R\, \psi_1(z_i)
\Big]\\[1ex]
&\quad= c_{N_c} \det_{fg}\Big[\bar\psi_f(z_i)\, \mathbb{P}_R\, \psi_g(z_i)\Big]\,.
\end{split}
\end{align}
Plugging this into \Eq{eq:ZQeffansatz}, we find
\begin{align}\label{eq:VeffLO}
V_\text{eff,LO}^{\qq+}\Big|_\text{singlet} = \frac{\kappa_{Q,\text{LO}}}{K_{Q,N_f}}\, \prod_{i=1}^Q \int\!d^4z_i\, \det_{fg}\Big[\bar\psi_f(z_i)\, \mathbb{P}_R\, \psi_g(z_i)\Big]\,,
\end{align}
with the coupling $\kappa_{Q,\text{LO}}$ given in \Eq{eq:kappaQLO}.
For any number of flavors, the color singlet channel is a determinant in flavor, since
the structure of the gauge group integration in \Eq{eq:uintnf2} is
\begin{align}
\int dU \prod_i U^{a_i b_i} (U^\dagger)^{c_i d_i} = c_{N_c N_f} \sum_{\sigma} \prod_i \delta_{a_i d_{\sigma(i)}} \delta_{b_i c_{\sigma(i)}} + (\text{non-singlet})\,,
\end{align}
where $\sigma(i)$ are permutations of $i = 1,\dots,N_f$, see {\it e.g.}\ \cite{Creutz:1978ub}.

To obtain the effective action we need to exponentiate $V_\text{eff,LO}^{\qq+}$.
So far we considered the generating functional in the background of a single $Q$-instanton in the limit of small constituent instantons.
For a single $Q$--anti-instanton we replace the right-handed with the left-handed projector
operator, $\mathbb{P}_R \rightarrow \mathbb{P}_L$, in \Eq{eq:VeffLO} to obtain
$V_\text{eff,LO}^{\qq-}$.
We now assume that the field configurations of topological charge $Q$ are described by a dilute gas of $Q$-instantons and anti-instantons. This generalizes the dilute instanton gas in \cite{tHooft:1986ooh} to arbitrary topological charge. 
For small constituent-instantons, the complete contribution of $Q$-instantons to the functional integral
of a dilute instanton gas is a simple statistical ensemble,
\begin{align}
\sum_{\nu_+=1}^\infty \sum_{\nu_-=1}^\infty \frac{(\kappa_{Q,\text{LO}}/K_{Q,N_f})^{\nu_+ + \nu_-}}{\nu_+! \, \nu_-!} 
\Big( V_\text{eff,LO}^{\qq+} \Big)^{\nu_+} \Big( V_\text{eff,LO}^{\qq-} \Big)^{\nu_-} = \exp\bigg[ \frac{\kappa_{Q,\text{LO}}}{K_{Q,N_f}} \Big( V_\text{eff,LO}^{\qq+} + V_\text{eff,LO}^{\qq-}\Big) \bigg]\, ;
\end{align}
$\nu_+$ and $\nu_-$ are the numbers of instantons and anti-instantons.
The anomalous contribution to the effective action is
\begin{align}
\begin{split}
\Delta S_\text{eff,LO}^{\qq} = -\int\!d^4x\, \frac{\kappa_{Q,\text{LO}}}{K_{Q,N_f}} \bigg\{&
\det_{fg}\Big[\bar\psi_f(x)\, \mathbb{P}_R\, \psi_g(x)\Big] \prod_{i=1}^{Q-1} \int\!d^4y_i \det_{fg}\Big[\bar\psi_f(y_i)\, \mathbb{P}_R\, \psi_g(y_i)\Big] \\[1ex]
&+
\det_{fg}\Big[\bar\psi_f(x)\, \mathbb{P}_L\, \psi_g(x)\Big] \prod_{i=1}^{Q-1} \int\!d^4y_i \det_{fg}\Big[\bar\psi_f(y_i)\, \mathbb{P}_L\, \psi_g(y_i)\Big] 
\bigg\}\,.
\end{split}
\end{align}
In principle, instantons of any topological charge contribute to the functional integral. Of course, in the semi-classical regime the contributions with higher topological charge are exponentially suppressed due to the factor $\exp(- 8 \pi^2 Q/g^2)$ in the instanton density. This picture is therefore not in conflict with lattice results on the topological charge at large temperature \cite{Bazavov:2012qja,Buchoff:2013nra,diCortona:2015ldu,Bonati:2015vqz,Borsanyi:2015cka,Brandt:2016daq,Dick:2015twa,Petreczky:2016vrs}. Still, these contributions can be present and the resulting anomalous contribution to the effective action of a dilute gas of instantons and anti-instantons of all topological charges is
\begin{align}
\Delta S_\text{eff} = \sum_Q \Delta S_\text{eff}^{(Q)}\,.
\end{align}
While the the effective interactions are certainly small in the dilute instanton gas, they might have relevant phenomenological implications as they manifest in anomalous correlation functions of higher order.

\subsection{The local interaction for $Q=2$}

We now repeat the analysis for $Q = 2$, taking into account the results of App.\ \ref{app:Q2Nf1} and \ref{app:Q2Nf}.
Instead of $\sim {\cal V}^2$ in the space-time volume ${\cal V}$, these are $\sim {\cal V}$.  We take
\begin{align}\label{eq:Z2effansatz}
\begin{split}
  Z_\text{eff}^{(2)+}[\bar J] &= \int\! \mathcal{D}\psi \mathcal{D}\bar\psi \exp\bigg\{\!-S[\chi]
  + \int_x \bar\psi \bar J \psi \bigg\} \,V_\text{eff}^{(2)+}\,,\\[2ex]
  V_\text{eff}^{(2)+} &=  \frac{\kappa_{2}}{K_{2,N_f}} \int\! d^4z\,
  \prod_{i=1}^{2} \Bigg\{\int\! dU_i \prod_{f=1}^{N_f} \big[\bar\psi_f(z)\,
  \omega_i \big] \big[\bar \omega_i\, \psi_f(z) \big] \Bigg\}\,.
\end{split}
\end{align}
Following the previous analysis, averaging over $V_\text{eff}^{(2)+}$ gives
\begin{align}\label{eq:Zeff2corr}
\begin{split}
  &\Bigg\langle\frac{1}{(2 N_f)!} \prod_{i=1}^{2} \prod_{f=1}^{N_f}\bigg[\int_{x_{fi}}
  \, \bar\psi(x_{fi}) \bar J(x_{fi}) \psi(x_{fi})\bigg]\,
 V_\text{eff}^{(2)+}\Bigg\rangle  \\[1ex]
&\quad = \kappa_{2}\, \int\!d^4z  \prod_{i=1}^2 \int\!dU_i\, \prod_{f=1}^{N_f} \int\!d^4x_{fi}\,
\bar\omega_i\, \Delta(x_{fi}-z)\, \bar J (x_{fi})\, \Delta(x_{fi}-z)\, \omega_i\,.
\end{split}
\end{align}
Choosing $\kappa_2$ and the tensors $\omega$ so that this correlation
function is identical to the generating functional for $Q = 2$ in \Eq{eq:nfZ2},
\begin{align}
\begin{split}
&\kappa_{2}\, \int\!d^4z  \prod_{i=1}^2 \int\!dU_i\, \prod_{f=1}^{N_f} \int\!d^4x_{fi}\,
\bar\omega_i\, \Delta(x_{fi}-z)\, \bar J (x_{fi})\, \Delta(x_{fi}-z)\, \omega_i\\[1ex]
&\quad= \big(8\pi^2\big)^{2 N_f} \int\! d\rho_2\, n_1(\rho_2)\rho_2^{N_f}
 \int\! d\rho_1\, n_1(\rho_1) \rho_1^{5 N_f} \,\mathcal{I}_{N_f,2} \\[1ex]
 &\qquad\times \int\!d^4 z \prod_{i=1}^2 \bigg\{\int\!dU_i \prod_{f=1}^{N_f} \int\! d^4x_{fi}\, U_i^\dagger \varphi_R^\dagger \Delta(x_{fi}-z) \bar{J}(x_{fi}) \Delta(x_{fi}-z) \varphi_R U_i  \bigg\}\, ,
\end{split}
\end{align}
which is valid for any $N_f$.
The overlap integral $\mathcal{I}_{1}$ is given by \Eq{eq:OI1} and
$\mathcal{I}_{N_f}$ for $N_f \geq 2$ by \Eq{eq:OINf}. From this we infer
\begin{align}\label{eq:kappa2loc}
\kappa_{2} = \big(8\pi^2\big)^{2 N_f}
  \int\! d\rho_2\, n_1(\rho_2)\rho_2^{N_f} \int\! d\rho_1\, n_1(\rho_1) \rho_1^{5 N_f}\, \mathcal{I}_{N_f,2}\,.
\end{align}
The determination of $\omega$ and the integration over the gauge group are as before.
The main difference here is that all propagators connect to the same point $z$.
For the channel which is a product of color singlet operators,
\begin{align}\label{eq:Veff2}
  V_\text{eff}^{(2)+}\Big|_\text{singlet} = \int_z \, \frac{\kappa_2}{K_{2,N_f}}
  \Big[\det_{fg}\Big(\bar\psi_f(z)\, \mathbb{P}_R\, \psi_g(z)\Big)\Big]^2\,.
\end{align}
A dilute gas of small $2$-instantons gives
\begin{align}
\Delta S_\text{eff}^{(2)} = - \int\!d^4x\, \frac{\kappa_2}{K_{2,N_f}} \bigg\{\Big[\det_{fg}\Big(\bar\psi_f(x)\, \mathbb{P}_R\, \psi_g(x)\Big)\Big]^2  
+ \Big[\det_{fg}\Big(\bar\psi_f(x)\, \mathbb{P}_L\, \psi_g(x)\Big)\Big]^2 \bigg\}\,,
\end{align}
with $\kappa_2$ in \Eq{eq:kappa2loc} and $K_{2,N_f} = 2^{N_f}/(2 N_f)!$.

\section{A low-energy model}\label{app:low}

We use our results to construct a linear sigma model (LSM) for two flavors.
For the sake of generality we include one more anomalous quartic term which is generated by instantons
with $Q=1$; we set it to zero in the main text.
The anomalous part of the effective Lagrangian is
\begin{align}
\mathcal{L}_{\mathcal{G}_\text{A}} &= -\chi_1 \big(\det\Phi + \det\Phi^\dagger\big)
 -\chi_2 \big[  \big(\det\Phi\big)^2 + \big(\det\Phi^\dagger\big)^2 \big]
 + \bar\lambda_3 \big(\Tr\, \Phi^\dagger\Phi\big)\big(\det\Phi + \det\Phi^\dagger\big)\,.
\end{align}
We set $\bar\lambda_3 = 0$ in the main text. The meson field is given by
\begin{align}\label{eq:phimes}
\Phi = (\sigma + i \eta) + (\vec{a}_0 + i \vec{\pi}) \vec{\tau}\,.
\end{align}
The equations of motion are
\begin{align}
\frac{\delta \Gamma}{\delta \phi}\bigg|_{\phi = \bar\phi} = 0\, ;
\end{align}
$\phi = (\sigma,\vec{a}_0,\eta,\vec{\pi})$, with
the vacuum expectation value $\bar\phi = (\bar\sigma,\vec{0},0,\vec{0})$,
\begin{align}\label{eq:sigmavev2}
\bar\sigma^2 = \frac{- 2 (m^2 - \chi_1)}{\lambda_1 + 2 \lambda_2 + 2\bar\lambda_3 - \chi_2}\,. 
\end{align}
When $m^2 > \chi_1 $ the expectation value of $\phi$ vanishes.
When $m^2 < \chi_1 $ $G_\text{qu}$ spontaneously breaks to $SU_V(2) \times \mathbb{Z}_2^A$.
If there were no anomalous terms, $\mathcal{L}_{{\cal G}_{\rm{qu}}} = 0$, $U_A(1)$ would also break,
resulting in four Goldstone bosons $\vec{\pi}$ and $\eta$. In the presence of the anomalous terms $U_A(1)$
is broken explicitly and only pions are massless. Due to isospin symmetry, there are four distinct masses,
\begin{align}\label{eq:symmasses2}
\begin{split}
m_\sigma^2 &= m^2 + \chi_1 - \frac{3}{2} (\lambda_1 + 2 \lambda_2 + 2 \bar\lambda_3  - \chi_2)\, \bar\sigma^2\,,\\
m_\pi^2 &= m^2 + \chi_1 - \frac{1}{2} (\lambda_1 + 2 \lambda_2 + 2 \bar\lambda_3  - \chi_2)\,\bar\sigma^2\,,\\
m_\eta^2 &= m^2 - \chi_1 - \frac{1}{2} (\lambda_1 + 2 \lambda_2 + 3 \chi_2)\,\bar\sigma^2\,,\\
m_{a_0}^2 &= m^2 - \chi_1 - \frac{1}{2} (3 \lambda_1 + 2 \lambda_2 + \chi_2)\,\bar\sigma^2\,.
\end{split}
\end{align}
In the symmetric phase, $\bar\sigma = 0$, only the quadratic terms contribute to the
masses directly and the $Q=1$ term $\chi_1$ induces a splitting of the
chiral pairs $(\sigma,\,\vec{\pi})$ and $(\eta,\,\vec{a}_0)$.
In the symmetric phases higher order couplings only influence these masses via loop corrections.
Inserting the expectation values from \Eq{eq:sigmavev2},
\begin{align}\label{eq:bromasses2}
\begin{split}
m_\pi^2 &= 0\,,\\
m_\sigma^2 &= -2 (m^2 - \chi_1)\,,\\
m_\eta^2 &= \frac{-2 m^2 (2\chi_2-\bar\lambda_3) + 2 \chi_1 (\lambda_1 + 2\lambda_2 + \bar\lambda_3 + \chi_2 )}{\lambda_1 + 2\lambda_2 + 2\bar\lambda_3 - \chi_2}\,,\\
m_{a_0}^2 &= \frac{-2 m^2 (\lambda_1 +\chi_2- \bar\lambda_3) + 2 \chi_1 (2\lambda_1 + 2\lambda_2 + \bar\lambda_3)}{\lambda_1 + 2\lambda_2 + 2 \bar\lambda_3 - \chi_2 }\,.
\end{split}
\end{align}
The pion is always a Goldstone boson.
To explore the influence of the anomalous terms, we 
fix the masses in vacuum using the following observables:
\begin{align}\label{eq:vacvals2}
\begin{split}
f_{\pi} = \bar\sigma_0 = 93\,\text{MeV}\,,\quad m_{\sigma,0} = 400\,\text{MeV}\,, \quad m_{\eta,0} = 820\,\text{MeV}\,,\quad m_{a_0,0} = 980\,\text{MeV}\,.
\end{split}
\end{align}
The $\eta$ mass is taken from Ref. \cite{Hashimoto:2008xg}.
For the other masses, we chose values compatible with Ref. \cite{Tanabashi:2018oca}.
Note that we identify the $\sigma$ meson with the $f_0(500)$. 
Within the mean-field approximation, and in absence of effects from topological charge $Q>1$,
taking $\bar\lambda_3 = 0$ all parameters, including $\chi_1$, are fixed by the vacuum masses.
This then also fixes the amount of axial symmetry breaking above the chiral phase transition,
as the only anomalous contribution to the masses in the symmetric phase stems from $\chi_1$.
When $\chi_2 \neq 0$ the vacuum mass spectrum can be fixed for different values for $\chi_2$,
and we can explore the influence of interactions induced by higher topological charge on the mass spectrum.

For a given value of $\chi_1$, the value of $m^2$ determines whether the symmetry is broken or not.
Thus in mean field theory varying temperature is equivalent to changing $m^2$.
We fix the masses in the vacuum according to \Eq{eq:vacvals2} and study the mass spectrum as a function of $m^2$
for different values of $\chi_2$.   We then find the following relations:
\begin{align}\label{eq:parasolfull}
\begin{split}
m^2 &= \frac{1}{2}(m_\eta^2 - m_\sigma^2)- \frac{\bar\sigma^2}{2} (2\chi_2 - \bar\lambda_3)\,,\\
\lambda_1 &= \frac{m_{a_0}^2 - m_\eta^2}{\bar\sigma^2}+ \chi_2\,,\\
\lambda_2 &= \frac{m_\sigma^2 + m_\eta^2 -m_{a_0}^2}{2 \bar\sigma^2} - \bar\lambda_3\,,\\
\chi_1 &= \frac{1}{2} m_\eta^2 -  \frac{\bar\sigma^2}{2} (2\chi_2 - \bar\lambda_3)\,.
\end{split}
\end{align}
Since we have six model parameters, but use only four parameters to fix them, we can
choose $\chi_2$ and $\bar\lambda_3$ as free parameters. 
\Eq{eq:parasolfull} implies that the anomalous quadratic coupling $\chi_1$ is determined by the anomalous quartic couplings $\chi_2$ and $\bar\lambda_3$ via
\begin{align}\label{eq:chi1fix}
\chi_1 &= \frac{1}{2} m_{\eta,0}^2 -  \frac{f_\pi^2}{2} (2\chi_2 - \bar\lambda_3)\,.
\end{align}
Setting $\chi_2 = \bar\lambda_3 = 0$, the coupling $\chi_1 = m_{\eta,0}^2/2$.
Conversely, if we set $\chi_1 = \bar\lambda_3 = 0$, then the coupling $\chi_2 = m_{\eta,0}^2/(2 f_\pi^2)$. If we use $\chi_1$ in the following, we mean $\chi_1(\chi_2,\bar\lambda_3)$ as defined by \Eq{eq:chi1fix}.

The system has two characteristic scales in $m^2$. The vacuum scale $m_\text{vac}^2$
is where the masses in the broken phase in \Eq{eq:bromasses2}
assume their vacuum values \eq{eq:vacvals2}. It can be read off from \Eq{eq:parasolfull},
\begin{align}
m_\text{vac}^2 = \chi_1 - \frac{1}{2}m_{\sigma,0}^2\,.
\end{align}
There is also the critical scale $m_\text{crit}^2$ for chiral symmetry breaking.
It is defined as the value of $m^2$ where the expectation value $\bar\sigma$ \eq{eq:sigmavev2} vanishes,
\begin{align}\label{eq:sigmavev2a}
m_\text{crit}^2 =  \chi_1 = m_\text{vac}^2 + \frac{1}{2} m_{\sigma,0}^2\,.
\end{align}
Hence, the characteristic scales of the system change as $\chi_2$ and $\bar\lambda_3$ vary.
To meaningfully compare masses for different values of $\chi_2$ and $\bar\lambda_3$, we define the reduced temperature
\begin{align}\label{eq:tred}
t = \frac{m^2-m_\text{vac}^2}{m_\text{crit}^2-m_\text{vac}^2}\,,
\end{align}
and rewrite the masses in terms of $t$,
\begin{align}
m^2(t) = m_\text{vac}^2 + t (m_\text{crit}^2-m_\text{vac}^2)\,.
\end{align}
$t = 0$ is the vacuum and $t = 1$ is where the phase transition occurs. As a function of $t$,
\begin{align}
\bar\sigma^2(t) = 
\begin{cases}
\,(1-t) f_\pi^2 & t \leq 1 \\
\, 0  & t > 1
\end{cases}
\end{align}
With this parametrization, the masses take very simple forms,
\begin{align}\label{eq:massesoft2}
\begin{split}
m_\sigma^2(t) &= 
\begin{cases}
\,(1-t)\, m_{\sigma,0}^2 & t \leq 1 \\
\,\frac{1}{2}(t-1)\, m_{\sigma,0}^2 & t > 1
\end{cases}\\[8pt]
m_\pi^2(t) &= 
\begin{cases}
\,0 & t \leq 1 \\
\,\frac{1}{2}(t-1)\, m_{\sigma,0}^2 & t > 1
\end{cases}\\[8pt]
m_\eta^2(t) &= 
\begin{cases}
\,m_{\eta,0}^2 - t\, (m_{\eta,0}^2 - 2 \chi_1) & t \leq 1 \\
\,\frac{1}{2}(t-1)\, m_{\sigma,0}^2 + 2 \chi_1 & t > 1
\end{cases}\\[8pt]
m_{a_0}^2(t) &= 
\begin{cases}
\,m_{a_0,0}^2 - t\, (m_{a_0,0}^2 - 2 \chi_1) & t \leq 1 \\
\, \frac{1}{2} (t-1)\, m_{\sigma,0}^2 + 2 \chi_1  & t > 1
\end{cases}
\end{split}
\end{align}
We see that $m_\sigma(t)$ and $m_\pi(t)$ are independent of the anomalous couplings. $\sigma$ is the critical mode that becomes massless at the phase transition. The mass splitting between the chiral pairs $(\sigma,\,\vec{\pi})$ and $(\eta,\,\vec{a}_0)$ in the symmetric phase is induced by $\chi_1$. This mass splitting vanishes in the limit $t\rightarrow\infty$. 
In absence of the quartic couplings, $\chi_2 = \bar\lambda_3 = 0$, one has $2 \chi_1 = m_{\eta,0}^2$ so the $\eta$ mass is independent of $t$ in the broken phase.
An interesting observation is that for $0 < (2\chi_2 - \bar\lambda_3) < m_{\eta,0}^2/f_\pi^2$, $m_\eta$ is a strictly decreasing function of $t$ in the broken phase and strictly increasing in the symmetric phase. Hence, it has a minimum at the chiral phase transition. This behavior can therefore be attributed to corrections related to couplings induced by topological charge two.

The masses and the characteristic scales $m_\text{vac}$ and $m_\text{crit}$ only depend on $\chi_1$. Hence, due to \Eq{eq:chi1fix}, only the combination of anomalous quartic couplings $(2\chi_2 - \bar\lambda_3)$ is relevant.
We want to focus on the coupling $\sim \chi_2$ which is induced by instantons with $Q = \pm 2$.
An analysis of the vacuum stability of the effective potential shows that
$\bar\lambda_3 \leq m_{\sigma,0}^2 / 4 f_\pi^2 \approx 4.62$.
We therefore conclude that setting $\bar\lambda_3 = 0$, which we made in the main text, is innocuous.
Most importantly, \Fig{fig:masses} does not change qualitatively for any value of $\bar\lambda_3$.
Our estimated values for the couplings $\chi_1$ and $\chi_2$ do change, however.
The larger $\bar\lambda_3 < 0$ is, the smaller $\chi_1$ and $\chi_2$ become.
Even so, $\bar \lambda_3$ has to become very large to have any significant effect.

By bosonizing the multi-quark interactions generated by $Q$-instantons,
the fermionic couplings $\kappa_Q$ can be related to the anomalous mesonic couplings $\chi_Q$.
For two flavors, $\kappa_1$ is a four-quark coupling and can readily be bosonized
by means of a Hubbard-Stratonovich transformation.
The 2-instanton term $\kappa_2$ is an 8-quark interaction for two flavors,
so more elaborate path integral bosonization techniques are required
\cite{Reinhardt:1988xu}. Here we adopt a simplistic bosonization scheme
motivated by low-energy models where mesons are coupled to quarks through Yukawa interactions, {\it i.e.}\
quark-meson models. Using the equations of motion, mesons are proportional to quark bilinears,
so we make a simple ansatz based upon \Eq{eq:phimes},
\begin{align}
\Phi = \frac{1}{2 M^2} \big[ (\bar\psi \psi + \bar\psi \gamma_5 \psi) + (\bar\psi \vec{\tau} \psi + \bar\psi \gamma_5 \vec{\tau} \psi)\vec{\tau} \big]\,.
\end{align}
$M$ is a fundamental parameter of our effective theory, and has dimensions of mass. By using the identity
\begin{align}
\epsilon^{ij} \epsilon^{fg} = \frac{1}{2}\Big[ \delta^{if} \delta^{jg} - (\tau^a)^{if} (\tau^a)^{jg}  \Big]\,,
\end{align}
the instanton-induced quark determinant becomes
\begin{align}
\begin{split}
\det(\bar\psi \mathbb{P}_{R} \psi) &= \frac{1}{2} \epsilon^{ij} \epsilon^{fg}\big(\bar\psi_i \,\mathbb{P}_{R}\, \psi_f\big)\big(\bar\psi_j \,\mathbb{P}_{R}\, \psi_g\big)\\[1ex]
&= \frac{1}{8} \Big[\big(\bar\psi \psi + \bar\psi \gamma_5\psi \big)^2  - \big(\bar\psi \vec{\tau}\psi + \bar\psi \gamma_5\vec{\tau} \psi \big)^2  \Big]\\[1ex]
&= \frac{M^4}{2} \det\Phi\,,
\end{split}
\end{align}
and similarly for the anti-instanton term,
\begin{align}
\det(\bar\psi \mathbb{P}_{L} \psi) =  \frac{M^4}{2} \det\Phi^\dagger \,.
\end{align}
The 1-instanton induced effective interaction transforms as
\begin{align}
\frac{\kappa_1}{K_{1,2}} \big[ \det(\bar\psi \mathbb{P}_{R} \psi) + \det(\bar\psi \mathbb{P}_{L} \psi)  \big] = \frac{\kappa_1 M^4}{2 K_{1,2}} \big( \det\Phi + \det\Phi^\dagger  \big),
\end{align}
and for the 2-instanton term,
\begin{align}
\frac{\kappa_2}{K_{2,2}} \Big\{\big[\det(\bar\psi \mathbb{P}_{R} \psi)\big]^2 + \big[\det(\bar\psi \mathbb{P}_{L} \psi) \big]^2\Big\} = \frac{\kappa_2 M^8}{4 K_{2,2}} \Big[\big(\det\Phi\big)^2 + \big(\det\Phi^\dagger\big)^2\Big].
\end{align}
We therefore identify
\begin{align}
\chi_1 = \frac{\kappa_1 M^4}{2 K_{1,2}} \,, \qquad \chi_2 = \frac{\kappa_2 M^8}{4 K_{2,2}}\,.
\end{align}
By plugging this into the expressions for the mesons masses above,
the dependence on $\chi_1$ and $\chi_2$ is replaced by a dependence only on $M$,
provided that we know the values of $\kappa_1$ and $\kappa_2$.
This reduces the number of independent parameters to four. Given the four input parameters \eq{eq:vacvals2},
at the level of mean field theory
the effective Lagrangian of \Eq{eq:DL} is uniquely determined.

\section{The instanton density}\label{app:diga}

In vacuum the instanton density is given by \Eq{instanton_density}, where
\begin{align}
d_{\overline{MS}} = \frac{2 e^{5/6}}{\pi^2 (N_c-1)! (N_c-2)!} e^{-1.511374 N_c + 0.291746 N_f}\, ;
\end{align}
$g^2(\rho \Lambda_{\overline{MS}})$ is the running coupling constant at two loop order,
\begin{equation}
  g^2(x)= \frac{(4 \pi)^2}{\beta_0 \log(x^{-2})} \left(1 - \frac{\beta_1}{\beta_0^2} \frac{\log(\log(x^{-2}))}{\log(x^{-2})} \right) \,,
  \end{equation}
with $\beta_0 = (11 N_c - 2 N_f)/3$ and
$\beta_1 = 34 N_c^2/3 - (13 N_c/3 - 1/N_c ) N_f$.
$\Lambda_{\overline{MS}}$ is the renormalization mass scale of QCD in the modified minimal subtraction scheme.
This expression is valid for small $x$, where $\log(x^{-2})$ is positive.
By asymptotic freedom,
the coupling $g^2(\rho \Lambda_{\overline{MS}})$ is small at small $\rho$, so instantons
are suppressed by the exponential of the
classical action, $8 \pi^2/g^2$.
Of necessity in a semi-classical computation, the exponential from the classical action
dominates over the prefactor, $\sim g^{-12}$, which arises from the Jacobian for
the collective coordinates of the instanton
\cite{tHooft:1976rip,*tHooft:1976snw,*tHooft:1986ooh}.  Conversely, when $\rho$ increases so
does the coupling $g^2(\rho \Lambda_{\overline{MS}})$.  The instanton density increases at first,
but eventually decreases, suppressed by the prefactor from the Jacobian.
The instanton density $n_1(\rho \Lambda_{\overline{MS}})$
is illustrated in Fig. (\ref{fig:instanton_density}); as seen there, there is a
natural maximum when $\rho \sim 0.50 \Lambda_{\overline{MS}}$ in the vacuum. 

\begin{figure}[t]
  \centerline{  \includegraphics[width=.4\textwidth]{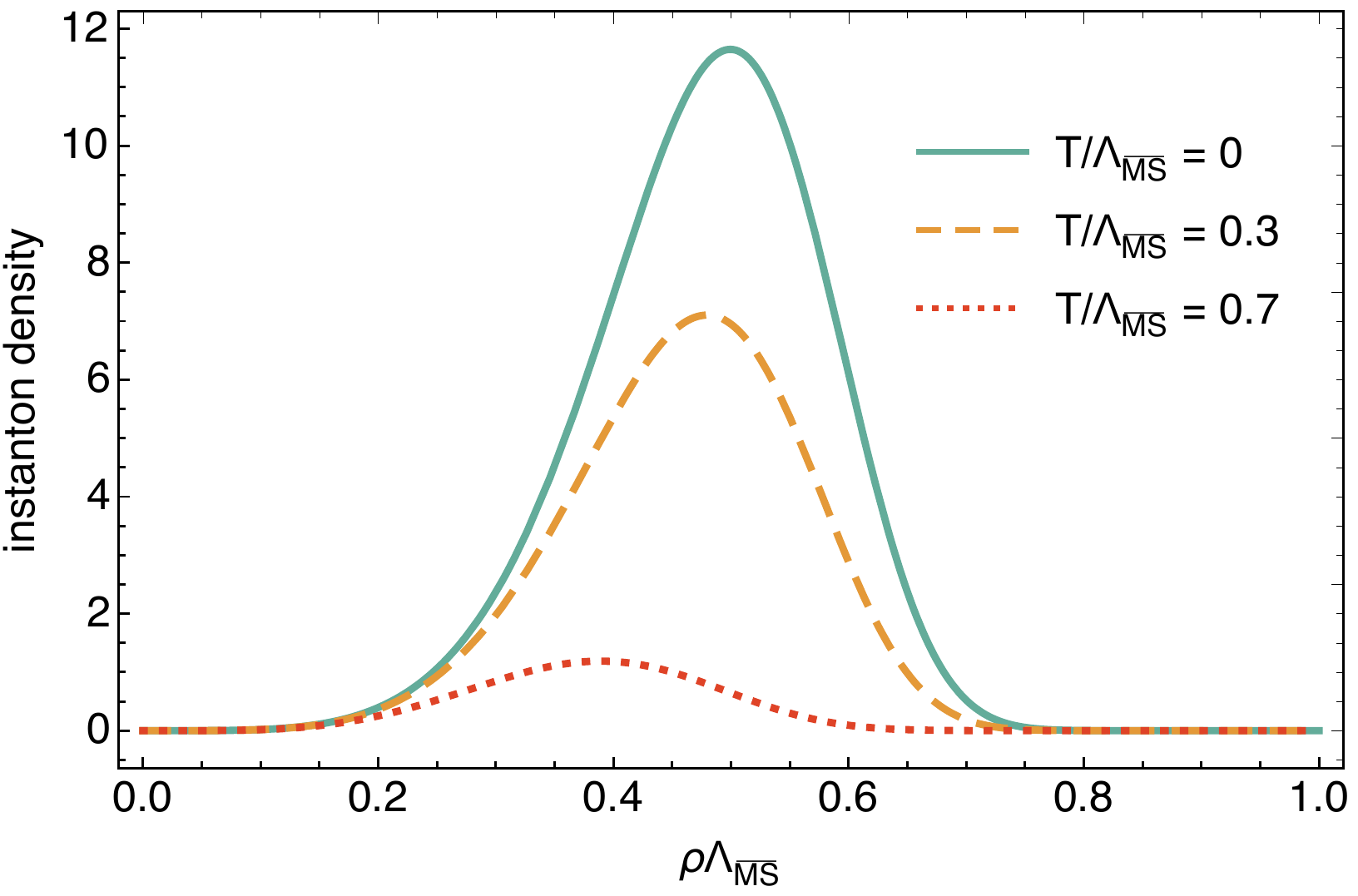} }
  \caption{The instanton density for a dilute instanton gas, Eq. \ref{instanton_density}, versus $\rho \Lambda_{\overline{MS}}$,
    for two massless quarks and three different temperatures at $\mu = 0$.}
  \label{fig:instanton_density}
\end{figure}

For a single instanton, at a temperature $T$ and quark chemical potential $\mu$,
we approximate the change to the instanton density as
\begin{equation}
  n_1(\rho,T,\mu) = \exp\bigg\{- \frac{2 \pi^2}{g^2} \rho^2 m_D^2 
    - 12 A(\pi \rho T) \bigg[1+ \frac{1}{6} (N_c - N_f)\bigg] \bigg\} \; n_1(\rho)\; ,
  \label{density_Tmu2}
\end{equation}
where
\begin{equation}
  m^2_D(T,\mu) = g^2 \left[\left(\frac{N_c}{3}+\frac{N_f}{6}\right)T^2 + \frac{N_f}{2 \pi^2}\, \mu^2 \right]\,,
    \label{debye_mass}
  \end{equation}
is the Debye mass at leading order, and \cite{Gross:1980br,Altes:2014bwa,*Altes:2015wla}
\begin{equation}
A(x) = - \frac{1}{12} \log\left(1 + \frac{x^2}{3}\right) + .0129
\left(1 + \frac{0.159}{x^{3/2}}\right)^{-8} \; .
\label{ax}
\end{equation}
The dominant term, $\sim \rho^2 m_D^2$, is straightforward to understand.
The topological charge is proportional to ${\rm tr} (\vec{E} \cdot \vec{B})$, where $\vec{E}$ and $\vec{B}$ are
the color electric and magnetic fields.  In any plasma, electrically charged particles screen static electric fields
over distances $\sim 1/m_D$.  Since instantons must carry color electric fields, 
by itself Debye screening suffices to suppress the instanton density.
Needless to say, this argument applies in a plasma where there is
Debye screning, and not at low temperature.

For a single instanton  at $T \neq 0$ and $\mu = 0$, to one loop order the instanton density can be computed analytically
either with puerile brute force \cite{Gross:1980br} or by being clever  \cite{Altes:2014bwa,*Altes:2015wla}.
The computation at $\mu \neq 0$ is,
unexpectedly, rather more difficult \cite{AragaodeCarvalho:1980de,*Baluni:1980db,*Chemtob:1980tu,Schafer:1998up,*Son:2001jm,*Schafer:2002ty,*Schafer:2002yy,*Toublan:2005tn,*Hatsuda:2006ps,*Yamamoto:2008zw,*Chen:2009gv,*Yamamoto:2009ey,*Brauner:2009gu,*Abuki:2010jq,*Fukushima:2010bq,*Eto:2011mk,*Mitter:2013fxa,*Mitter:2013fxa}, and
as of yet has not been computed for arbitrary values of $\rho \mu$.
At nonzero $\mu$, then, we only include the leading contribution of quarks to the Debye mass.
At $\mu = 0$ and $T \neq 0$, though, numerically one can show 
that for the instanton density, the relative
difference between the complete result and that with just the leading term from the Debye mass is small,
at most a few percent for all values of $\rho T$.  We comment that the instanton
density to one loop order could be computed at $\mu \neq 0$ 
numerically using the Gelfand-Yaglom method \cite{Coleman:1978ae}, as has been done
for the computation of the one loop determinant in an instanton field
for quarks of nonzero mass \cite{Dunne:2004sx,*Dunne:2005te,*Dunne:2005cy}.

Using the elementary ansatzes of Eqs. (\ref{instanton_density}), (\ref{density_Tmu1}) and (\ref{density_Tmu2}),
we can calculate numerically
how the density changes with temperature and chemical potential.  Consider first $T \neq 0$ and $\mu = 0$.
As illustrated in the left plot of Fig. (\ref{fig:temp}), as the Debye mass increases the instanton density decreases smoothly.
To have some definite measure, we
define the temperature as that where the integrated instanton density is $1/10^{\rm th}$ its value
at zero temperature as $T_I$.  
For three colors and two massless flavors, $T_I^{2fl} \approx 0.71 \Lambda_{\overline{MS}}$; 
for three massless flavors, $T_I^{3fl} \approx 0.74 \Lambda_{\overline{MS}}$.
Using $\Lambda_{\overline{MS}} \approx 332$~MeV \cite{Tanabashi:2018oca},
for two flavors $T_I^{2fl} \approx 236$~MeV, and $T_I^{3fl} \approx 246$~MeV for three.

\begin{figure}[t]
  \centerline{  
  \includegraphics[width=.4\textwidth]{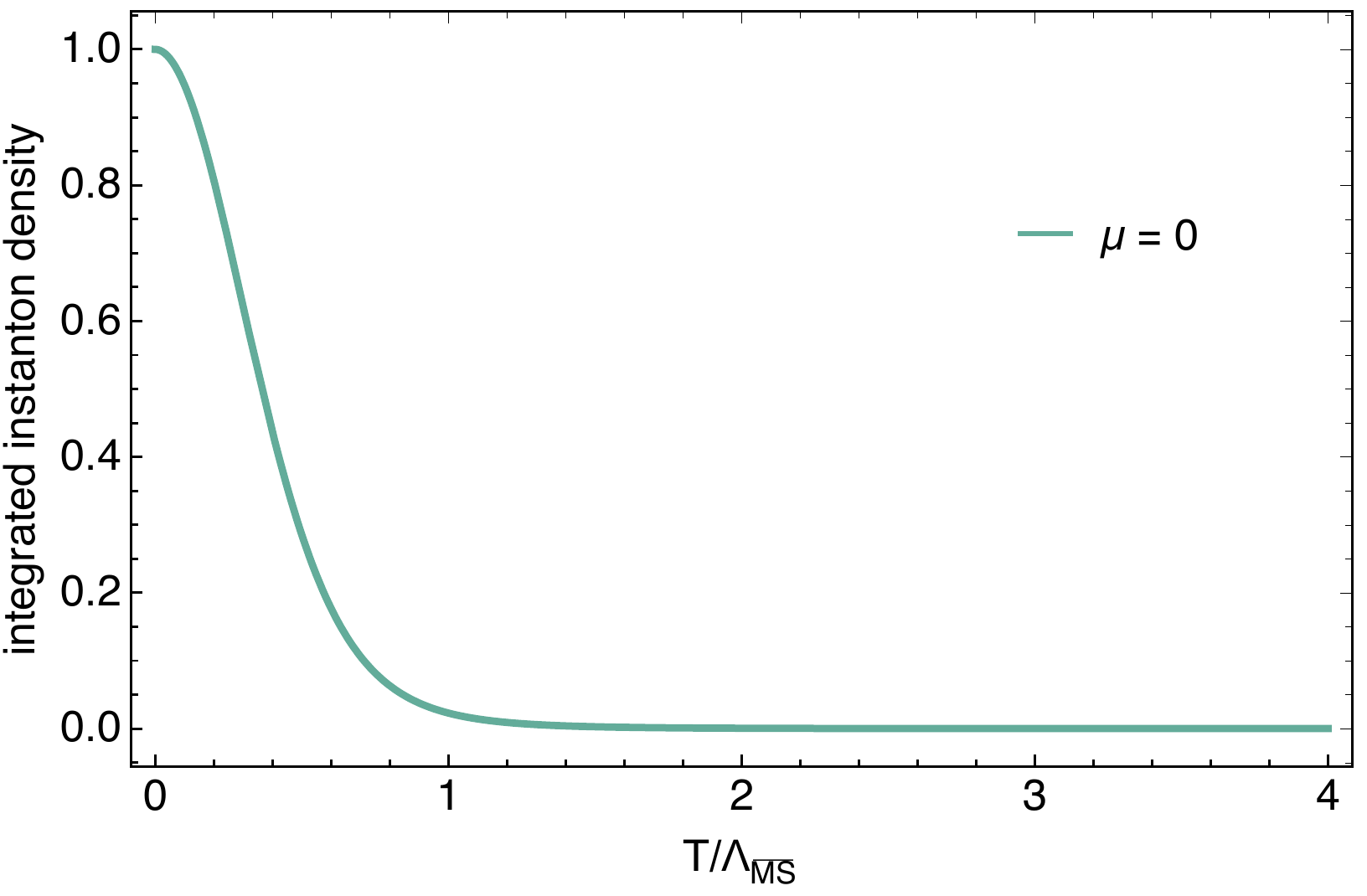}
  \hspace{7ex}
  \includegraphics[width=.4\textwidth]{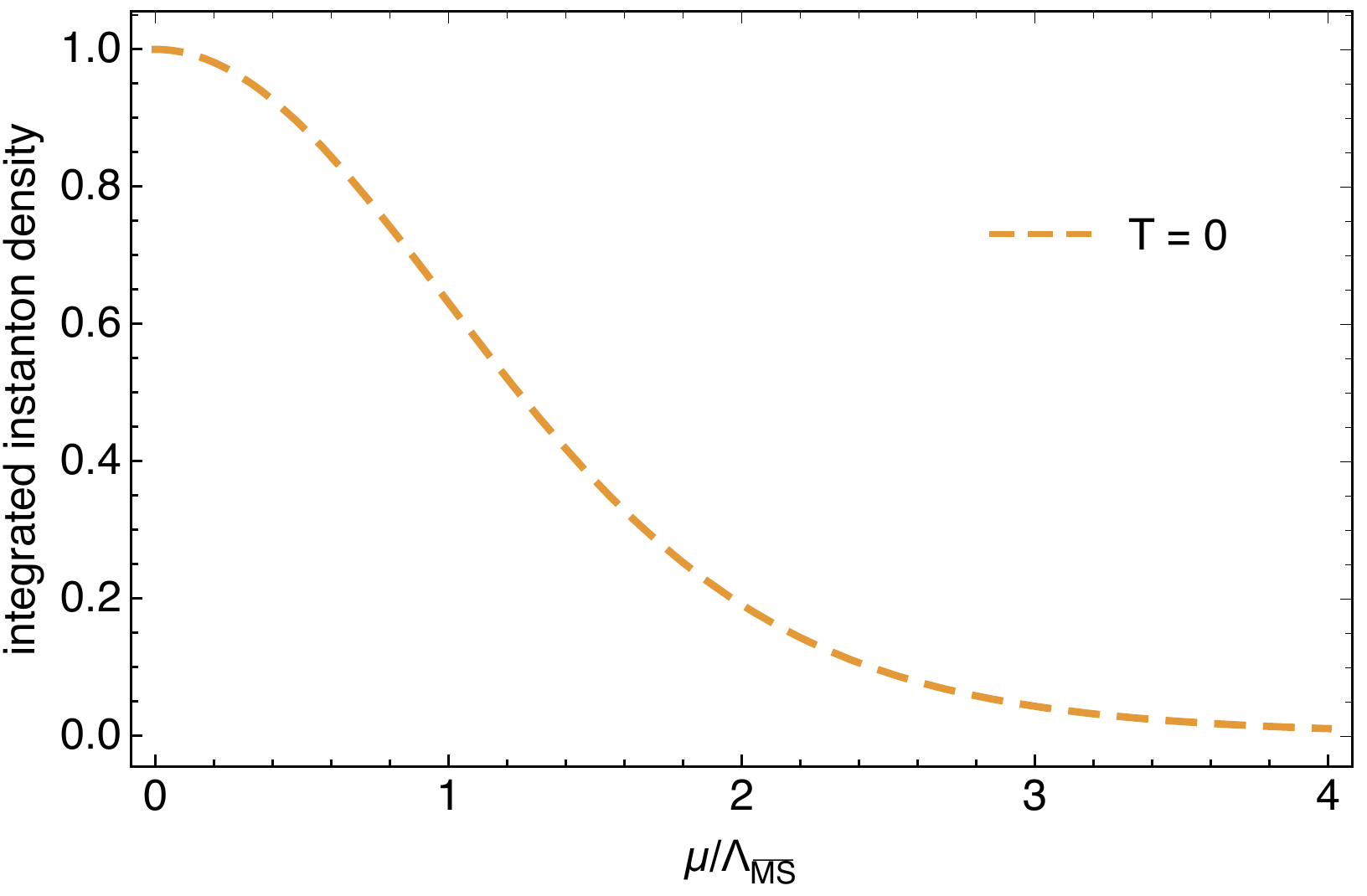} 
  }
  \caption{For two flavors, the ratio of the density of an instanton gas, integrated over $\rho$
    and normalized to the vacuum value. The left plot shows this as function of
    $T/\Lambda_{\overline{MS}}$ at $\mu = 0$ and the right plot as a function of $\mu/\Lambda_{\overline{MS}}$ at $T = 0$.
    This demonstrates graphically that instantons evaporate {\it much} more slowly as the quark chemical potential
  increases, as opposed to increasing the temperature.}
  \label{fig:temp}
\end{figure}

We stress that these numerical values are, at best, merely suggestive.
Under our naive ansatz for a dilute instanton gas, the instanton density is very sensitive to the
choice of $\Lambda_{\overline{MS}}$; after all,
merely on dimensional grounds the instanton density is $\sim (\Lambda_{\overline{MS}})^4$.

At nonzero temperature, to date the results from lattice QCD
find that  above temperatures $300-400$~MeV, the fall off with temperature is a power law,
whose value follows from the classical action for a single instanton and the running of the coupling $g^2$
with temperature.  The overall prefactor measured in lattice QCD is approximately
ten times larger than the one loop result, but at high temperature perhaps this is ameliorated by the complete
computation at two loop order \cite{Morris:1984zi,*Ringwald:1999ze}.  It is still an open
question as to whether 
topologically non-trivial fluctuations become dilute
below \cite{Aoki:2012yj,*Cossu:2013uua,*Fukaya:2015ara,*Tomiya:2016jwr,*Aoki:2017paw,Brandt:2016daq} or above
\cite{Bazavov:2012qja,*Buchoff:2013nra,*Dick:2015twa,*Petreczky:2016vrs}
the appropriate transition temperature.  This is presumably due to a combination of
effects from fractional dyons and instantons with integral topological charge, either as a liquid
or a gas.  For our purposes, which is frankly phenomenological, the moral which we draw is that 
a dilute instanton gas is not a preposterous assumption, at least at temperatures about $T_\chi$.

Consider next the case of zero temperature and nonzero quark chemical potential.
As for temperature, the density of instantons are smoothly suppressed as $\mu$ increases.
The integrated density of instantons, shown in the right plot of \Fig{fig:temp}, is $1/10^{\rm th}$ that in vacuum when
$\mu_I^{2fl} \approx 2.44 \Lambda_{\overline{MS}}$ for two flavors, and
$\mu_I^{3fl} \approx 2.22 \Lambda_{\overline{MS}}$ for three flavors.
These correspond to $\mu_I^{2fl} \approx 810$~MeV for two flavors,
and $\mu_I^{3fl} \approx 737$~MeV for three.
Taking $T_\chi \approx 156$~MeV
\cite{Aoki:2006we,*Aoki:2006br,*Aoki:2009sc,*Bazavov:2011nk,*Borsanyi:2012ve,*Bazavov:2014pvz,*Bhattacharya:2014ara},
this is approximately $\sim 1.5 \, \pi T_\chi$,

While even the instanton density at one loop order is incomplete at $\mu \neq 0$, we note that these are
{\it extremely} high values of the quark chemical potential.  
They are almost into the perturbative regime, for $\mu > 1$~GeV \cite{Kurkela:2009gj,*Ghisoiu:2016swa,*Gorda:2018gpy}.

This gross disparity has a simple origin, and thus may persist a more careful analysis.  
In a thermal bath, or the Fermi sea of cold, dense quarks, instantons are suppressed primarily
because of Debye screening.  As can be seen from the expression for the
Debye mass in Eq. \ref{debye_mass}, the natural scale for the chemical potential is $\mu \approx \pi T$.
Indeed, as the Euclidean energy of any fermion field is an odd multiple of $\pi T$, this
balance between $\mu$ and $\pi T$ is true of the propagator at tree level.

The weak dependence upon the quark chemical potential can also be understood
in the limit of large $N_c$.  As $N_c \rightarrow \infty$ the coupling $g^2 \sim 1/N_c$,
so that if the number of quark flavors $N_f$ is held fixed as $N_c \rightarrow \infty$, any effects of quarks are
suppressed by $\sim 1/N_c$.  In the plane of $T$ and $\mu$, large $N_c$ then generates a ``quarkyonic'' regime
\cite{McLerran:2007qj,*McLerran:2008ua,*Andronic:2009gj,*Kojo:2009ha,*Kojo:2011cn,*Pisarski:2018bct,*McLerran:2018hbz,*Pisarski:2019cvo}.  Our naive estimate for a dilute instanton gas is simply another illustration of this.

At present, numerical simulations of lattice QCD with classical computers can only
provide results at nonzero temperature and $\mu \leq T$.
Simulations of cold, dense quark matter may be possible with quantum computers, but will not be available
for some time. Studying the dense regime of QCD with functional continuum methods,
on the other hand, is possible even in the near-term
\cite{Braun:2014ata, *Cyrol:2017ewj, *Fischer:2018sdj, *Fu:2019hdw}.
Still, using an effective model, such as a dilute gas of instantons,
is useful for developing the physical picture before results from first principles are available.

\end{widetext}
\bibliography{instanton}

\end{document}